\documentclass[twocolumn,3p]{elsarticle}

\usepackage{amssymb}
\usepackage{subfigure}
\usepackage{graphicx}
\usepackage{color}

\usepackage{lineno}

\usepackage[figuresright]{rotating}
\usepackage{upgreek}

\newcommand{\tp}[1]{\texttt{#1}}
\newcommand{\mod}{}
\newcommand{\modakira}{}
\newcommand{\rev}{}

\begin{document}

\begin{frontmatter}




\title{TARGET~5: a new multi-channel digitizer with triggering capabilities for gamma-ray atmospheric Cherenkov telescopes}


\author[SLAC,LANL]{A.~Albert}
\author[ECAP,SLAC]{S.~Funk}
\author[Nagoya]{T.~Kawashima}
\author[SLAC]{M.~Murphy}
\author[Nagoya,MPIK,Leicester]{A.~Okumura\corref{cor1}}
\ead{oxon@mac.com}
\author[SLAC]{R.~Quagliani}
\author[SLAC]{L.~Sapozhnikov}
\author[Nagoya]{H.~Tajima}
\author[MPIK,SLAC]{L.~Tibaldo\corref{cor1}}
\ead{luigi.tibaldo@mpi-hd.mpg.de}
\author[Wisconsin]{J.~Vandenbroucke\corref{cor1}}
\ead{justin.vandenbroucke@wisc.edu}
\author[Hawaii]{G.~Varner}
\author[Wisconsin]{T.~Wu}

\address[SLAC]{Kavli Institute for Particle Astrophysics and Cosmology, Department of Physics and SLAC National Accelerator Laboratory, Stanford University, Stanford, CA 94305, USA}
\address[LANL]{Physics Division, Los Alamos National Laboratory, Los Alamos, NM 87545, USA}
\address[ECAP]{Erlangen Centre for Astroparticle Physics, D-91058 Erlangen, Germany}
\address[Nagoya]{Institute for Space-Earth Environmental Research, Nagoya University, Furo-cho, Chikusa-ku, Nagoya, Aichi 464-8601, Japan}
\address[MPIK]{Max-Planck-Institut f\"{u}r Kernphysik, P.O. Box 103980, D 69029 Heidelberg, Germany}
\address[Leicester]{Formerly of Department of Physics and Astronomy, University of Leicester, Leicester, LE1 7RH, UK}
\address[Wisconsin]{Physics Department and Wisconsin IceCube Particle Astrophysics Center, University of Wisconsin, Madison, WI 53706, USA}
\address[Hawaii]{Department of Physics and Astronomy, University of Hawaii, 2505 Correa Road, Honolulu, HI 96822, USA}

\cortext[cor1]{Corresponding authors:}

\begin{abstract}
TARGET~5 is a new application-specific integrated circuit (ASIC) of the TARGET family, designed for the readout of signals from photosensors in the cameras of imaging atmospheric Cherenkov telescopes (IACTs) for ground-based gamma-ray astronomy. TARGET~5 combines sampling and digitization on 16 signal channels with the formation of trigger signals based on the analog sum of groups of four channels. {\rev We describe the ASIC architecture and performance.} TARGET~5 improves over the performance of the first-generation TARGET ASIC, achieving: tunable sampling frequency from {\rev $<0.4$~GSa/s to {\mod $>1$~GSa/s}}; a dynamic range on the data path of 1.2 V with {\mod effective dynamic range of 11}~bits and DC noise of ${\sim}0.6$~mV; 3-dB bandwidth of 500 MHz; {\rev crosstalk between adjacent channels $<1.3\%$};  {\mod charge resolution improving from 40\% to $<4\%$ between 3 photoelectrons (p.e.) and $>100$~p.e.} (assuming 4 mV per p.e.); and minimum stable trigger threshold of 20 mV (5 p.e.) with trigger noise of 5 mV (1.2 p.e.), {\rev which is} mostly limited by {\mod interference between trigger and sampling operations}. {\mod TARGET~5 is the first ASIC of the TARGET family} used in {\modakira an} IACT {\mod prototype}, {\rev providing one development path for readout electronics in the forthcoming Cherenkov Telescope Array (CTA)}.
\end{abstract}


\end{frontmatter}


\section{Introduction}
In the past three decades, imaging atmospheric Cherenkov telescopes (IACTs) have greatly advanced observations of very-high-energy gamma-ray emission from the Universe, with numerous implications for astrophysics, particle physics, and cosmology \cite[e.g.,][]{2009ARA&A..47..523H}. This field is {\mod soon going to be revolutionized} with the advent of the Cherenkov Telescope Array (CTA) \cite{2011ExA....32..193A}, which is going to increase the source sensitivity by an order of magnitude at energies from 100~GeV to 10~TeV and to extend observations to the ranges well below 100 GeV and above 100 TeV. 

{\rev The performance requirements of CTA drive innovation to improve performance and lower cost.  One innovative design is the Schwarzschild--Couder telescope \cite{2007APh....28...10V}, which features dual-mirror optics for excellent optical performance (focusing of Cherenkov photons) and a reduced camera plate scale compared to the traditional single-mirror (Davies-Cotton) design used so far for IACTs.  The reduced camera size enables compact, inexpensive, densely pixelated photodetectors such as silicon photomultipliers.  The optical performance combined with dense pixellation provides improved field of view, angular resolution, and hadronic background rejection capabilities~\cite{Wood201611}.}

TARGET is an application-specific integrated circuit (ASIC) series that has been designed for the processing of the photodetector signals in such telescopes. {\mod The goals in the inception of TARGET were to keep the costs low and integrate several functionalities in a compact design.} We have described the concept of TARGET~1, the first generation of ASICs of the TARGET family, and characterized its performance in \cite{2012APh....36..156B}.  Several improvements drove the development of TARGET~5 (after a few design iterations), which is described in this paper.

{\rev TARGET~5 is the first chip
of the TARGET family to be used in a telescope prototype, namely a prototype {\mod of} the Gamma-ray Cherenkov Telescope (GCT) \cite{2015arXiv150806472M}, a Schwarzschild--Couder small-sized telescope proposed in the framework of the CTA
project. TARGET~5 is used in the first prototype of the GCT camera, {\modakira also known as Compact High Energy Camera with
MAPMTs (CHEC-M)} \cite{2013arXiv1307.2807D,2015arXiv150901480D}. TARGET is also planned to be used in a medium-sized telescope proposed in the framework of the CTA project, namely the Schwarzschild-Couder Telescope (SCT)  \cite{SCTICRC2015}.}

Key features of TARGET are:
\begin{itemize}
\item a compact design that combines signal sampling and digitization, as well as triggering, for 16 channels in a single chip, which lowers the cost\footnote{${\sim}\$35$ per channel for the realization discussed in this paper, estimated $<$~\$20 per channel in a large production on the scale required for several dozen CTA telescopes.}, {\mod improves on reliability further reducing maintenance costs}, and enables the use with compact photodetectors such as multi-anode photomultiplier tubes (MAPMTs) or silicon photomultipliers in a compact camera design
\item a sampling frequency tunable up to $>1$~GSa/s, ideally suited
  for the measurement of the $\gtrsim 5$~ns pulses from Cherenkov
  flashes
\item a deep buffer (16,384 samples in TARGET~5) for large trigger latency tolerance between distant (${\sim}1$~km) telescopes\footnote{\rev Within CTA, using a hardware coincidence trigger between telescopes is not planned for GCTs, but it is foreseen for SCTs.}
\item dynamic range $>10$~bits
\item {\rev moderate power consumption, for the applications described in this paper $\lesssim20$~mW per channel}
\end{itemize}

{\mod TARGET ASICs are implemented for IACTs into} front-end electronics modules that combine all the functions described above to read out 64 photodetector pixels using six or fewer printed circuit boards, four ASICs, and a companion field-programmable gate array (FPGA) \cite{2012APh....36..156B}. The low number of components supports affordability and reliability.

The structure of this paper is as follows. Section~\ref{sec:architecture} describes the architecture
of the TARGET~5 ASIC. Section~\ref{sec:performance} presents the
characterization of its performance, including sampling and digitization in \ref{sec:datapath}, as
well as triggering in~\ref{sec:trigpath}. Section~\ref{sec:cameramod} briefly outlines how TARGET~5 is implemented into front-end electronics modules for CHEC-M, and Section~\ref{sec:conclusions} presents the conclusions and outlook.  

\section{The TARGET~5 architecture}\label{sec:architecture}

\begin{figure*}[htbp]
\includegraphics[width=1\textwidth]{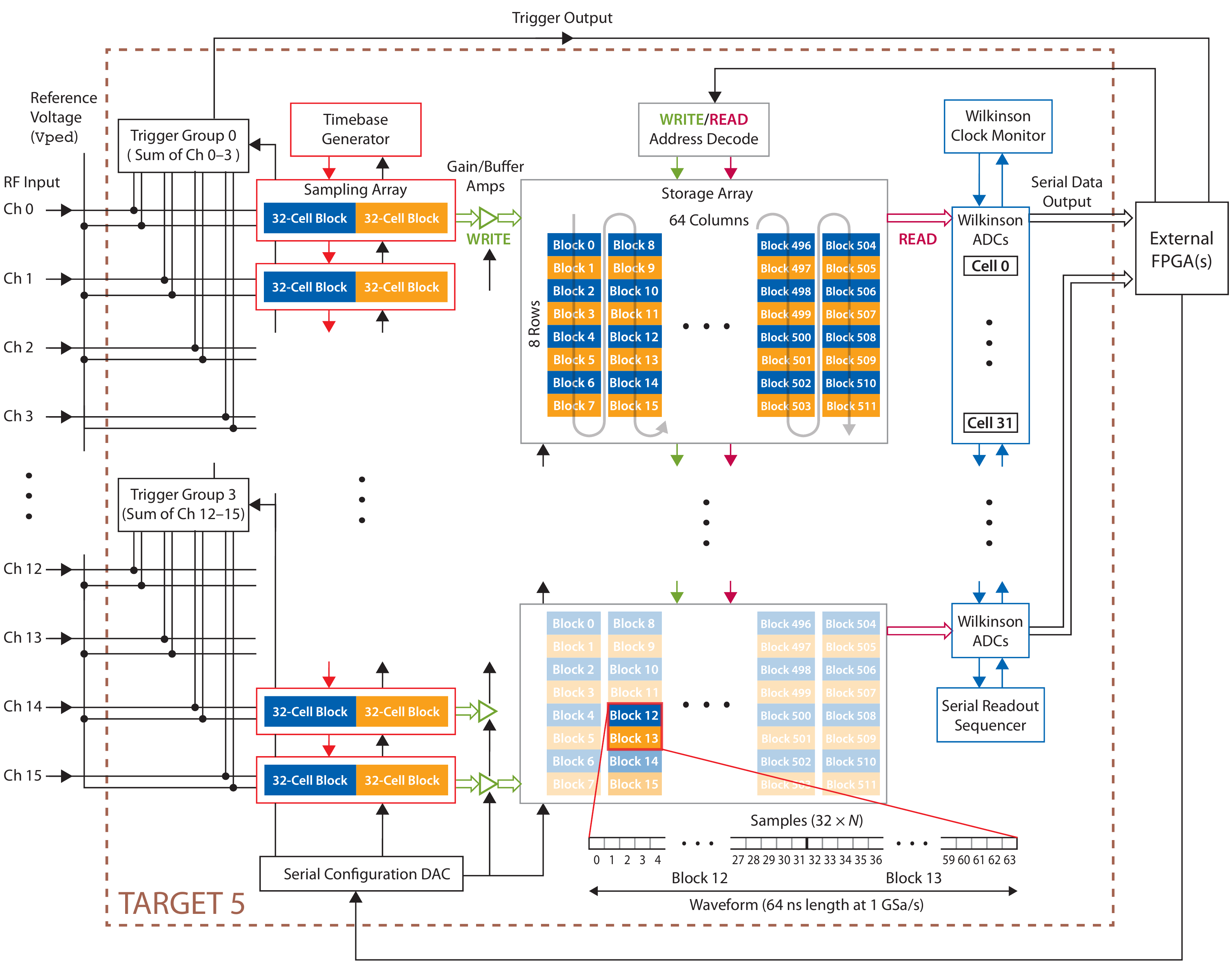}
\caption{Functional block diagram of the TARGET~5 ASIC, with key
  components shown.  Sixteen channels are processed both for trigger
  formation and analog sampling.
  The trigger generation is based on the analog sum of signals in four adjacent channels, and the trigger output is transmitted off-ASIC to higher-level logic that uses it to generate readout requests sent back to the ASIC.
  A {\mod timebase generator}
  controls the sampling of signals into two groups of 32 sampling
  capacitors per channel.  Ping-pong operation enables transfer of {\mod analog} samples from one group of 32
  to storage while the other group is sampling, with the roles
  reversed in the subsequent half sampling cycle. {\mod Each storage buffer is an array of 512 blocks (8 rows by 64 columns) with a total of 16,384 capacitors per channel.
  Blocks of 32}
  storage capacitors can be randomly accessed for digitization by onboard
  Wilkinson ADCs.  Sufficient  ADCs are included to digitize 32 samples per channel, and all 16 channels, in parallel.
  Individual converted samples may then be selected and serially
  transmitted off-ASIC on all 16 channels concurrently. {\modakira In this example, two blocks, corresponding to 64 samples (64 ns at 1 GSa/s), are read out as a digitized waveform.} Configuration
  of operating parameters, such as DACs for bias and control, are
  programmed through a serial-parallel interface.
}
\label{T5blockdiagram}
\end{figure*}

Fig.~\ref{T5blockdiagram} presents an overview of the major functional
blocks of the TARGET~5 ASIC: triggering, analog sampling and storage,
analog-to-digital converters (ADC), and configuration of
control features and digital-to-analog converters (DAC) provided via a
serial-parallel interface.
The general features of TARGET~5 detailed in the following text are summarized in Table~\ref{spectable} with comparison to TARGET~1.
\begin{table*}[htbp]
\centering
\begin{tabular}{lcc}
\hline
 & \textbf{TARGET~1} & \textbf{TARGET~5}\\
\hline
{\rev N}umber of channels & 16 & 16 \\
{\rev D}ynamic range of digitizer (bits) & 9 or 10 & 12 \\
{\rev S}ampling frequency (GSa/s) & $0.7$ to $2.3$ &  {\rev $<0.4$ to $>1$}\\
3-dB analog bandwidth (MHz) & 150 & 500\\
{\rev C}rosstalk at 3~dB frequency & $<4\%$ & $<2\%$ \\
{\rev S}ize of storage array (cells per channel) & 4,096 & 16,384 \\
Wilkinson ADC counter speed (MHz) & 445 & ${\sim}700$ \\
{\rev M}inimum digitization block (number of cells) & 16 & 32 \\
{\rev D}igitization time per block ($\upmu$s) & 2.3 (10 bit) & 5.9 (12 bit)\\
{\rev N}umber of Wilkinson ADCs & 32 & 512 \\
{\rev N}umber of cells digitized simultaneously & $16 \times 2$~channels & $32 \times 16$~channels\\
{\rev C}lock speed for serial data transfer (Mbps) & -- & 109 \\
{\rev C}hannels for simultaneous data transfer & -- & 16 \\
{\rev T}rigger outputs & 1 (OR of 16 channels) & 4 (analog sum of 4 channels)\\
\hline
\end{tabular}
\caption{Features of TARGET~5 compared to TARGET~1.}\label{spectable}
\end{table*}

As with TARGET 1 and other predecessors in the TARGET family, TARGET~5 is a 16-channel
device where both a signal and its reference signal (an input
pedestal voltage, \tp{Vped}) are input to the ASIC, to provide a
modest amount of common-mode noise rejection, as well as reference for
the trigger gain path. The input signal is simultaneously processed for sampling (data path) and triggering (trigger path).
  
{\mod ASICs of the TARGET family {\rev use} switched capacitor arrays (SCAs) to sample signals at very high sampling rate, i.e., the input is connected to an  array of capacitors via analog switches which are sequentially connected and disconnected to sample the signal at regular intervals.

{\rev The sampling buffer depth ($\sim$16~$\upmu$s) needed to trigger using coincidence between distant (up to 1~km) telescopes requires several thousand storage capacitors given a sampling frequency of 1~GSa/s. Directly driving the capacitance of such an array limits its analog bandwidth. Therefore, one of the major improvements of TARGET~5 over TARGET~1 is the separation of  sampling operations into two stages in order to simultaneously achieve large bandwidth and a deep buffer for trigger decisions. In the first stage, an SCA with a small number of cells (two blocks of 32 cells), called the sampling array, is used for signal sampling in order to reduce the capacitance load on the input. In the second stage, the samples are transferred to an SCA with a large number of cells (16,348), called the storage array, which provides the desired buffer depth.
While acquisition
occurs in one group of 32 cells in the sampling array, the other is written to the storage array.  This ping-pong approach provides continuous sampling.}}

Control of the sample timing is provided by a {\mod timebase generator} that
is driven by two digital signals sent from the FPGA to the ASIC: {\mod \tp{SSTin} and \tp{SSPin}}. These signals go
through {\mod time delay elements} that are current-starved inverters and control {\mod charge tracking and hold in the SCAs}. The
sampling speed of TARGET~5 is controlled {\mod by adjusting the supply current for the inverter in the delay elements}. 


{\mod Previous measurements of our timebase generator indicated that the sampling speed} is
temperature dependent with a coefficient of approximately 0.2\% per
$^\circ\mathrm{C}$ \cite{Varner:2007zz}. In order to reuce this effect,
there are two mechanisms available {\mod in TARGET~5}. The ASIC is equipped with a
continuous ring oscillator copy of the timebase generator (with one
additional inverter) and its output is available for external monitoring and feedback control in firmware/software.  {\mod The delayed copy of \tp{SSTin} after the full chain of time delay elements, \tp{SSTout}}, is also available for monitoring and feedback.

Blocks of 32 {\mod storage cells} are randomly accessible for readout from the
storage array on demand. Once selected, the 32
storage cells in all 16 channels are powered up for Wilkinson ADCs. A
Wilkinson ramp {\mod voltage} generator block generates and broadcasts a ramp with
adjustable duration and slew rate to all channels. The Wilkinson ramp
slew rate is adjusted by varying the capacitor charging current,
denoted \tp{Isel}, or by changing an external ramping capacitor.


At a separately controllable start time, a 12-bit ripple counter (with
adjustable speed) is begun for each channel. In order to support
the fastest possible digitization, separate oscillators are provided
for each counter.  When the voltage ramp crosses the comparator
threshold for a given sample {\mod voltage}, the counter stops and the count then
represents the time (ADC code) corresponding to the voltage held in
the storage cell. In order to maintain a constant Wilkinson clock rate
as a function of temperature, a separate, identical Wilkinson counter
is provided inside the Wilkinson clock block for monitoring {\mod and feedback}. Address decoding
and sequencing is performed inside a serial readout sequencer
block. Digitized {\mod values are randomly accessible for serial transfer} on all 16 channels in parallel.

{\rev Digitization and readout can occur on demand, initiated by an external trigger signal.  This external trigger signal can be generated by external {\rev logic} based on the trigger primitives generated by TARGET~5 itself.
{\rev Trigger primitives are formed based on the analog sum} of four adjacent channels, referred to as
a trigger group. {\rev Each ASIC provides four trigger primitives from four independent trigger groups.} To compensate for channel-to-channel photodetector {\rev gain variations}, bias voltages are provided to tune the gain of each channel (before the analog sum) between 1 and 6.5. Furthermore, the contribution from each individual
channel to the sum can be disabled in order to mask out noisy channels. The summed signal is routed to a
comparator for thresholding.}

\section{Performance}\label{sec:performance}

\subsection{Evaluation board and software}
\begin{figure}
\includegraphics[width=0.5\textwidth]{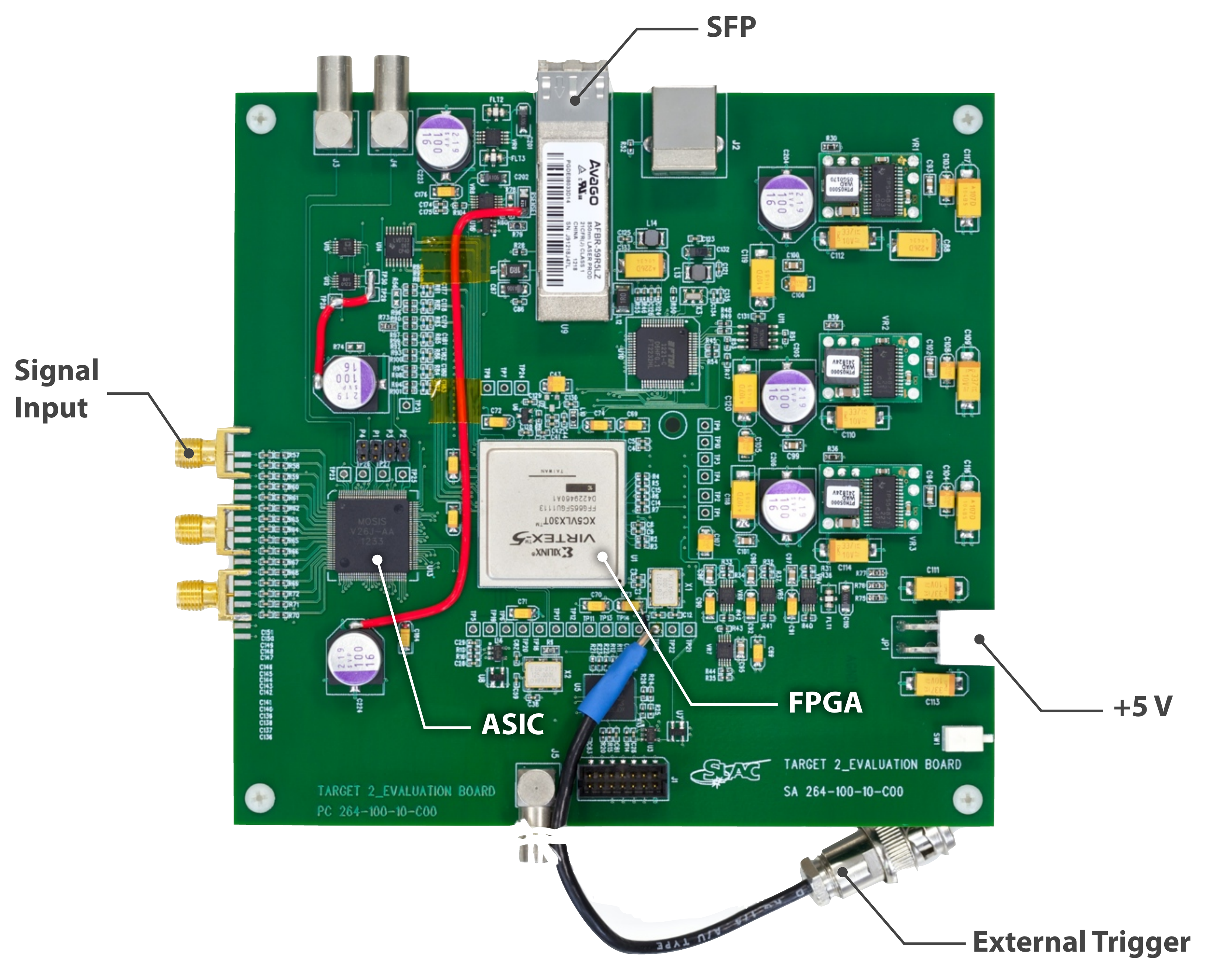}
\caption{TARGET~5 evaluation board. Some key elements are indicated: SMA connectors to input external signals, a coaxial cable to input external trigger signals, the $+5$~V power supply connector, and an SFP optical fiber connector that interfaces with an external computer.}
\label{fig:EvalBoardPhoto}
\end{figure}

Evaluation of TARGET~5 performance has been carried out using the dedicated board shown in Figure~\ref{fig:EvalBoardPhoto}. It features a TARGET~5 ASIC and an FPGA (Xilinx Virtex-5) as well as all the ancillary components necessary to support them. The FPGA controls the ASIC and connects to an external computer through a fiber interface. {\rev External termination resistors are used in order to set the real part of the input impedance to 50~$\Omega$ for interfacing with standard cables (the same strategy is adopted for the front-end electronics camera module described in Section~\ref{sec:cameramod}).}

{\mod The external computer and the evaluation board FPGA communicate with each other through an optical fiber/Gigabit ethernet cable via the User Datagram Protocol (UDP).}
The control and data acquisition software, \textit{libTARGET}, is written in C++11 to support both POSIX and Windows systems, and supports the generation of a Python wrapper. The latest version of \textit{libTARGET} can be obtained from the authors upon request.

\subsection{Data path}\label{sec:datapath}

\subsubsection{Sampling}


The analog sampling frequency is determined by two control voltages,
\tp{VdlyP} and \tp{VdlyN}{\mod, which set the supply and sink current for the time delay elements,  respectively}.  {\mod Each} control
voltage is supplied by a digital-to-analog converter (DAC) {\mod internal to the ASIC controlled through the serial-parallel interface}.  

\begin{figure}
\includegraphics[width=0.5\textwidth]{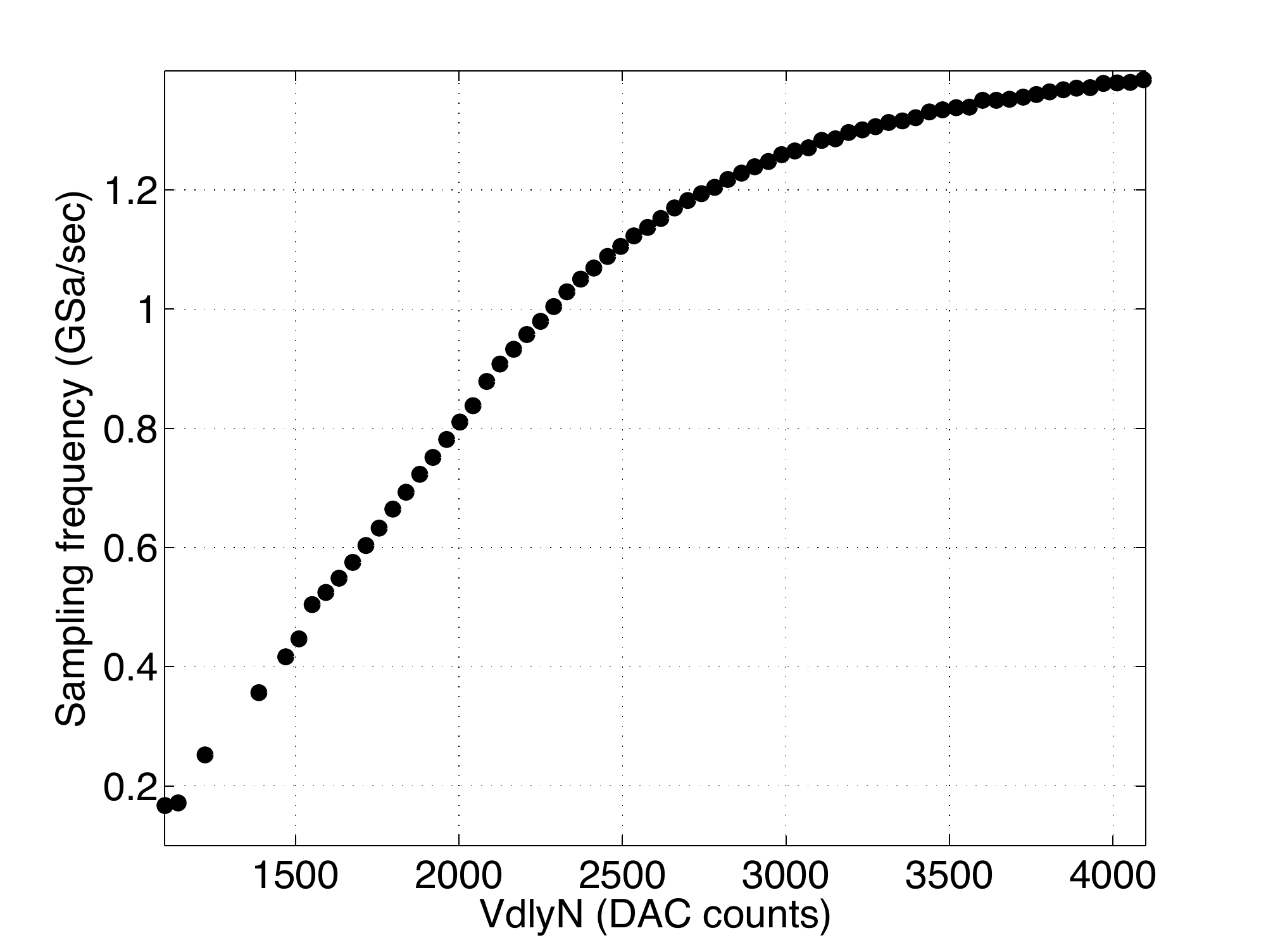}
\caption{Measured sampling frequency as a function of control voltage
  \tp{VdlyN}.}
\label{sampling-frequency}
\end{figure}

The measured sampling frequency as a function of \tp{VdlyN} is shown
in Figure~\ref{sampling-frequency}.  This measurement was made at room
temperature, with \tp{VdlyP} set to 1616 DAC counts.  The sampling
frequency was measured by recording a sinusoid of known frequency from
a function generator and fitting to determine the sampling frequency.
In general, frequencies from 0.2 to 1.4 GSa/s are possible.
However, the FPGA firmware must support a particular frequency in order to
use it.  To maintain phase {\rev alignment} while wrapping around the primary
sampling buffer of 64 samples, 64 samples must be a multiple of 8~ns
(one tick of the 125 MHz FPGA clock).  This limits the available
frequencies to a discrete set: 1.33~GSa/s, 1.14~GSa/s,
1.00~GSa/s, etc.  The frequencies that were explicitly supported {\rev in the FPGA firmware} and
tested are {\rev 1~GSa/s and 0.4~GSa/s}. {\rev Measurements presented in this paper were performed with a sampling frequency of 1~GSa/s.}

\begin{figure}
\includegraphics[width=0.5\textwidth]{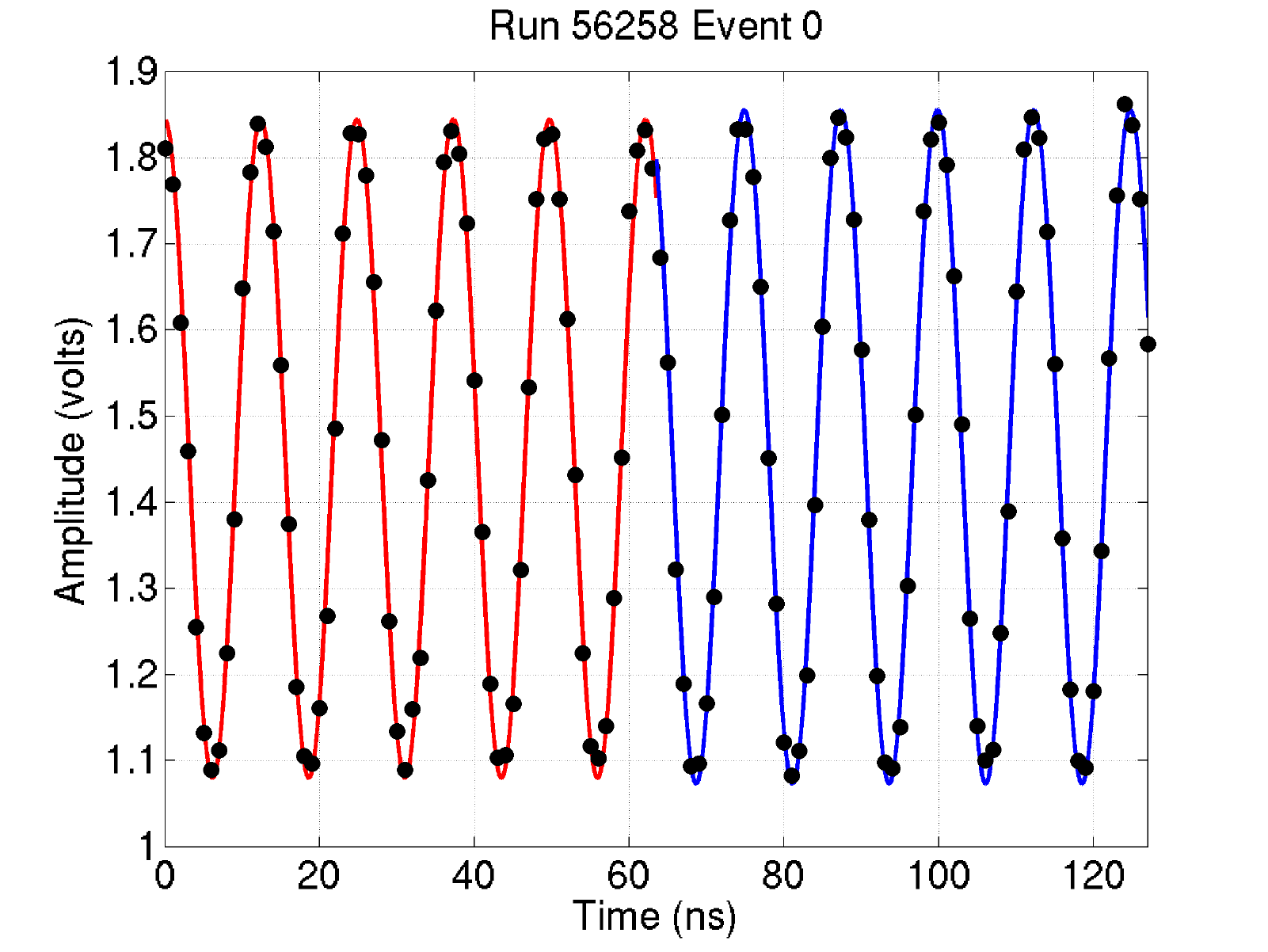}
\includegraphics[width=0.5\textwidth]{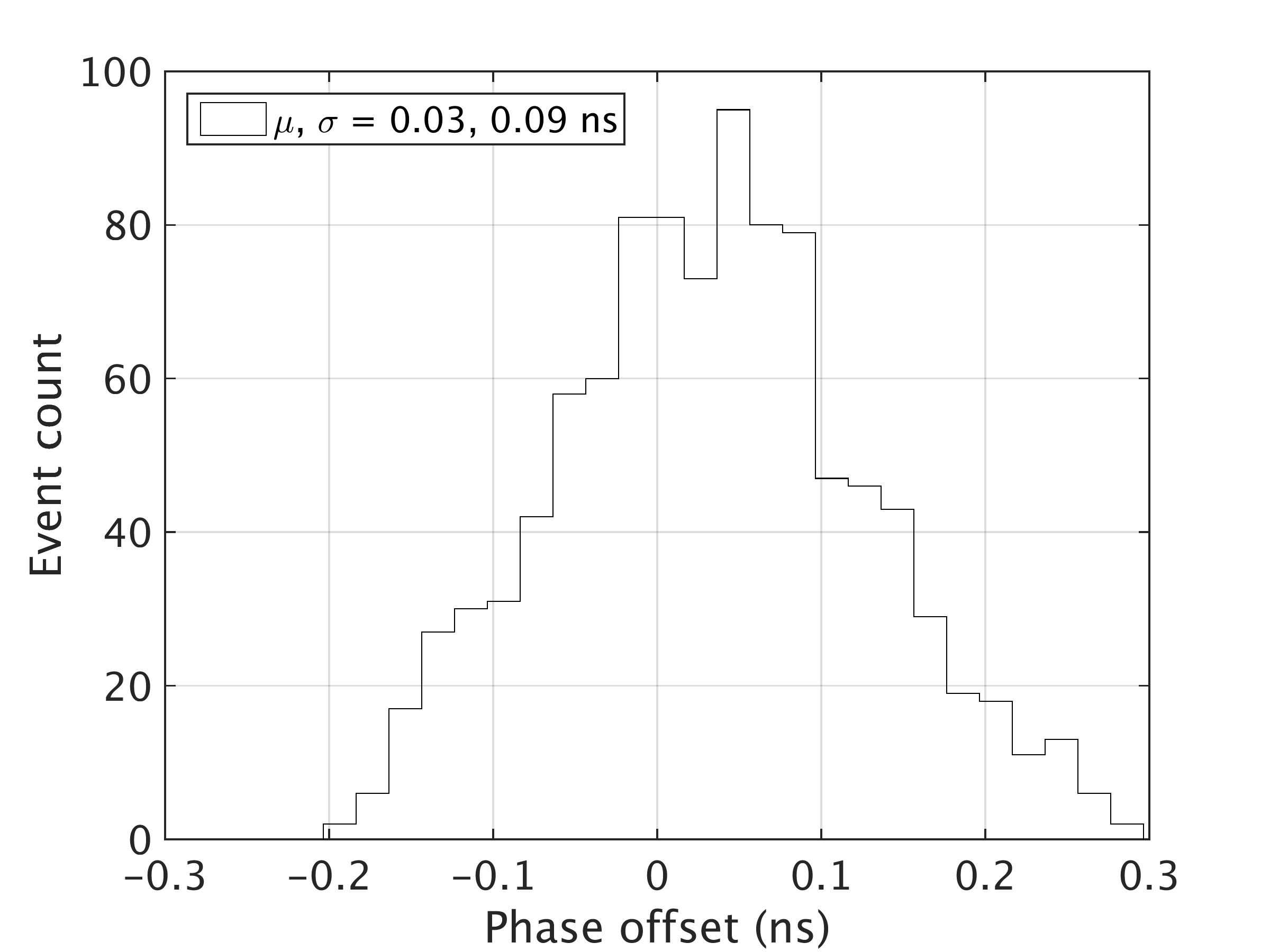}
\includegraphics[width=0.5\textwidth]{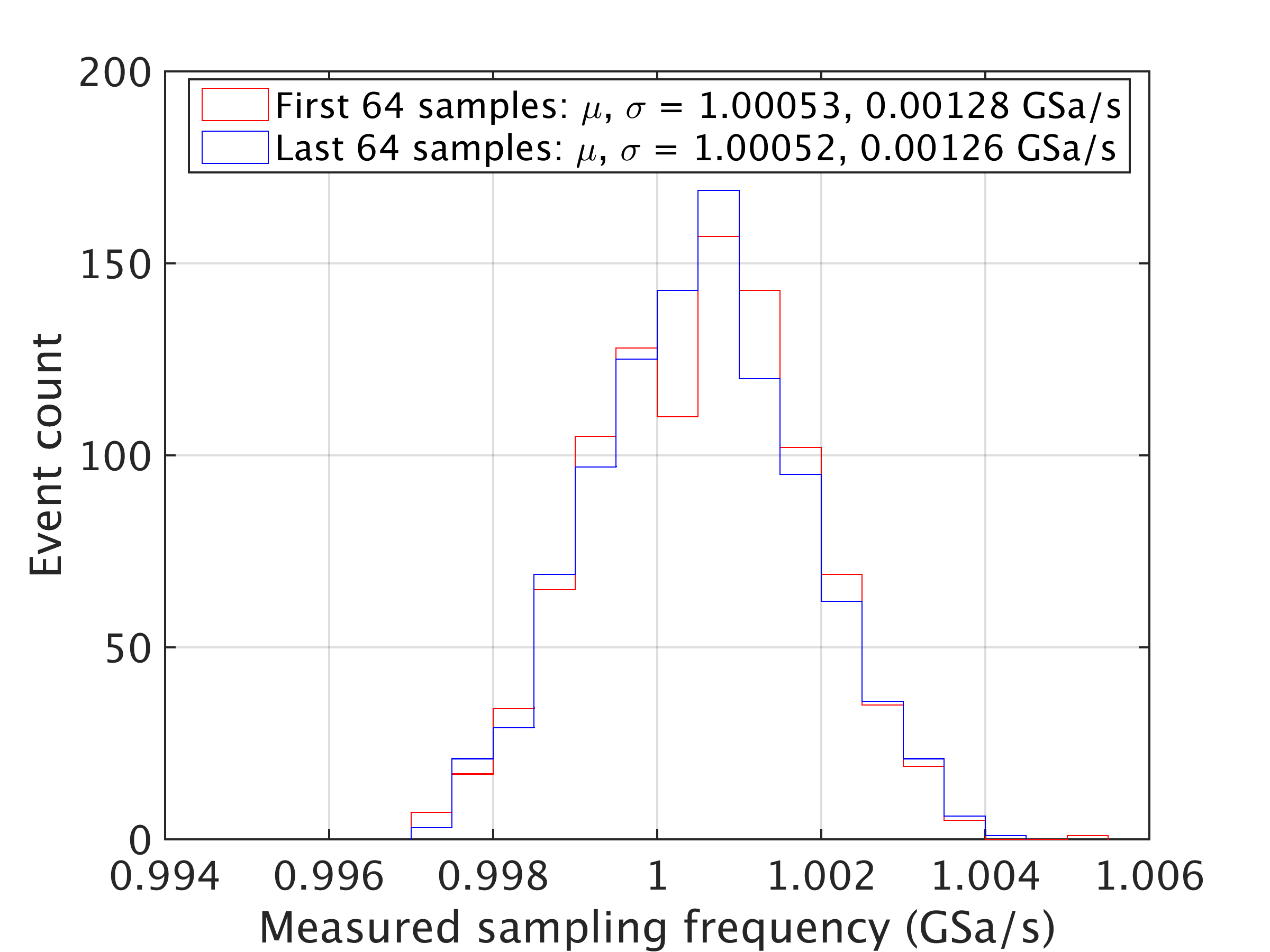}
\caption{\rev Top: 80~MHz sinusoid recorded with TARGET~5 evaluation board at 1~GSa/sec.  The first 64 samples correspond to one pass through the primary (sampling) buffer, and the last 64 samples correspond to a second pass.  Data points show the recorded samples and curves show fit results.  The two waveform halves are fit independently. Middle: histogram of fitted phase offset between the two waveform halves, for 997 such events.  Bottom: sampling frequency measured with the same data set.}\label{phase-lock}
\end{figure}

Precise timing is necessary to maintain a constant {\rev sampling frequency}, particularly when wrapping around the primary sampling buffer.  For fixed \tp{VdlyN}, the sampling frequency varies with temperature.  A feedback algorithm is used to control \tp{VdlyN} in order to achieve a stable sampling frequency despite temperature variation.  Several feedback mechanisms (implemented on the companion FPGA) were evaluated using a thermal chamber to vary the temperature between $-20$~$^\circ$C and $+50$~$^\circ$C. {\rev  An example waveform illustrating the achieved sampling frequency precision, as well as histograms of the wrap-around timing alignment and measured sampling frequency, are shown in Figure~\ref{phase-lock}.    In this case a digital clock manager in the FPGA is used for feedback. The phase of 
\tp{SSTout} (delayed copy of one of the input timing signals, \tp{SSTin}, after all the time delay elements) is stabilized against drift due to temperature
variation and other causes by the feedback loop controlling \tp{VdlyN}.  Feedback loop parameters were optimized based on measurements of sinusoids in the thermal chamber. This feedback loop achieves good temperature stability, corresponding to a phase gap of 0.1~ns introduced between succeeding blocks by a temperature change of 10$^\circ$~C.  Alternative feedback algorithms achieve slightly smaller temperature dependence and slightly larger event-to-event variation in sampling frequency.}

\begin{figure}
\includegraphics[width=0.5\textwidth]{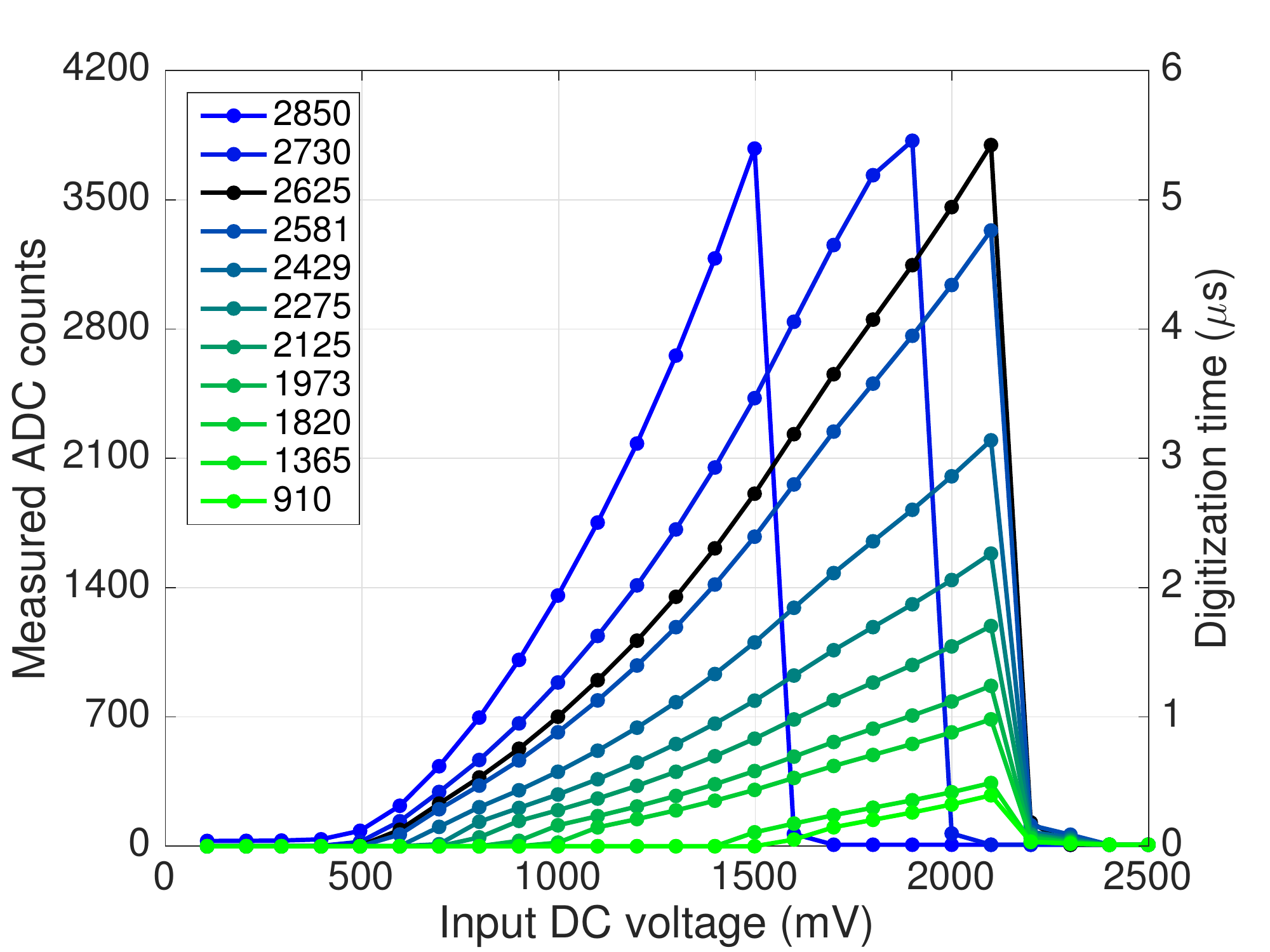}
\caption{Measured transfer function for a range of \tp{Isel} values
  (in DAC counts).  The default value used for most of the studies
  reported here is 2625~DAC counts.  The user can select a particular
  value of \tp{Isel} to choose a particular tradeoff between dynamic
  range, resolution, and dead time.  The digitization time assumes a 700~MHz Wilkinson clock.}\label{Isel-scan}
\end{figure}

\subsubsection{DC transfer function and noise}

The DC transfer function (measured ADC counts as a function of input signal amplitude) depends on several configuration parameters. 
The most important is \tp{Isel}, which {\rev controls the ramp} capacitor charging current\footnote{\rev Specifically, \tp{Isel} controls the bias voltage for the current mirror that generates the charging current. Larger  \tp{Isel} values correspond to smaller currents.}{\rev, hence} the slew rate of the Wilkinson ramp.  {\mod \tp{Isel} is set by an internal DAC.} The dependence of the transfer function and of the digitization time on \tp{Isel} is shown in Figure~\ref{Isel-scan}.  For a given Wilkinson clock frequency and DC signal amplitude, increasing \tp{Isel} slows the ramp, increasing the digitization time but also {\mod improving the voltage} resolution.  For a given application, \tp{Isel} can be selected to match the required input range and also to choose the tradeoff between digitization time, resolution, and noise.

\begin{figure}
\includegraphics[width=0.5\textwidth]{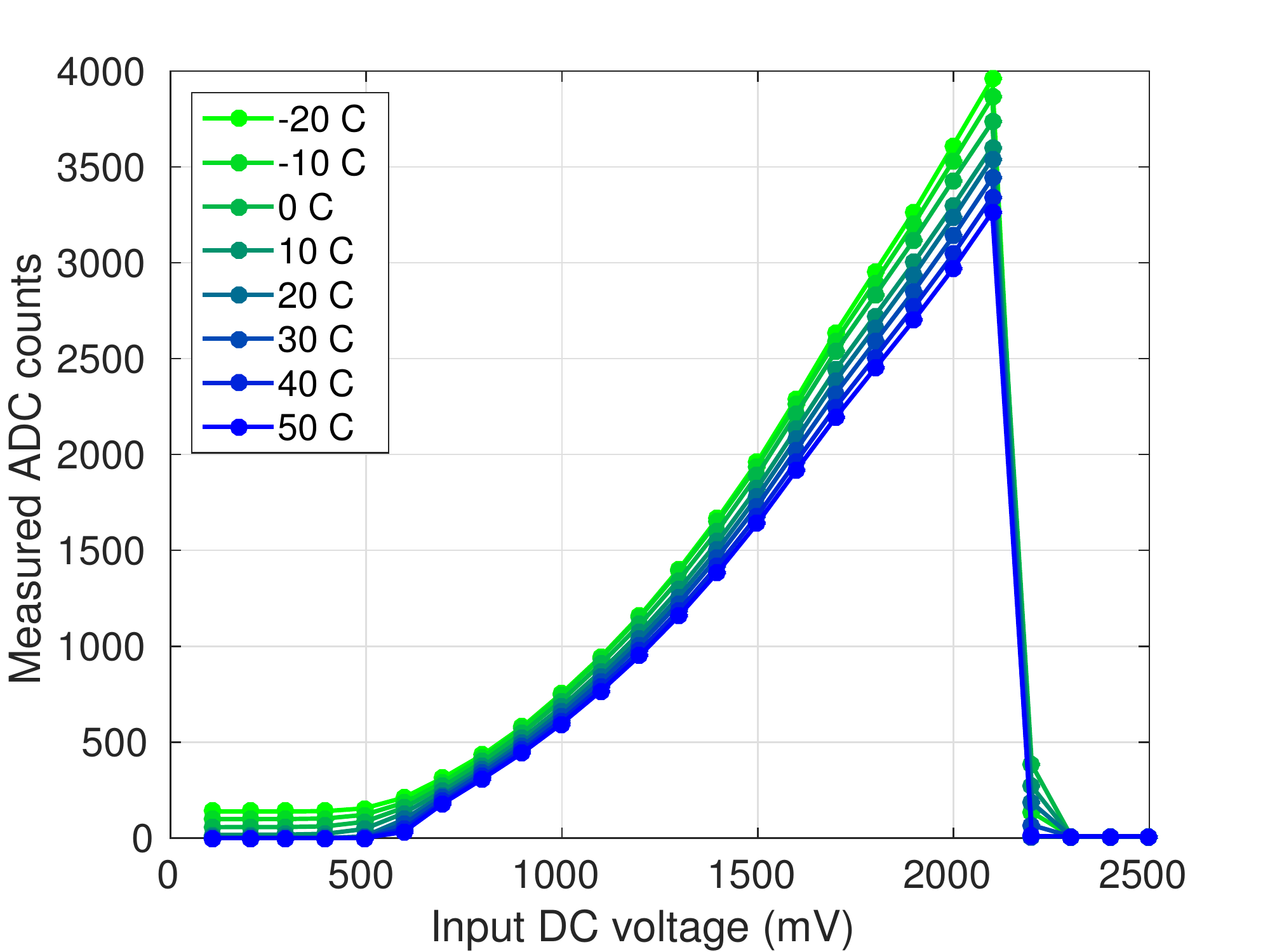}
\includegraphics[width=0.5\textwidth]{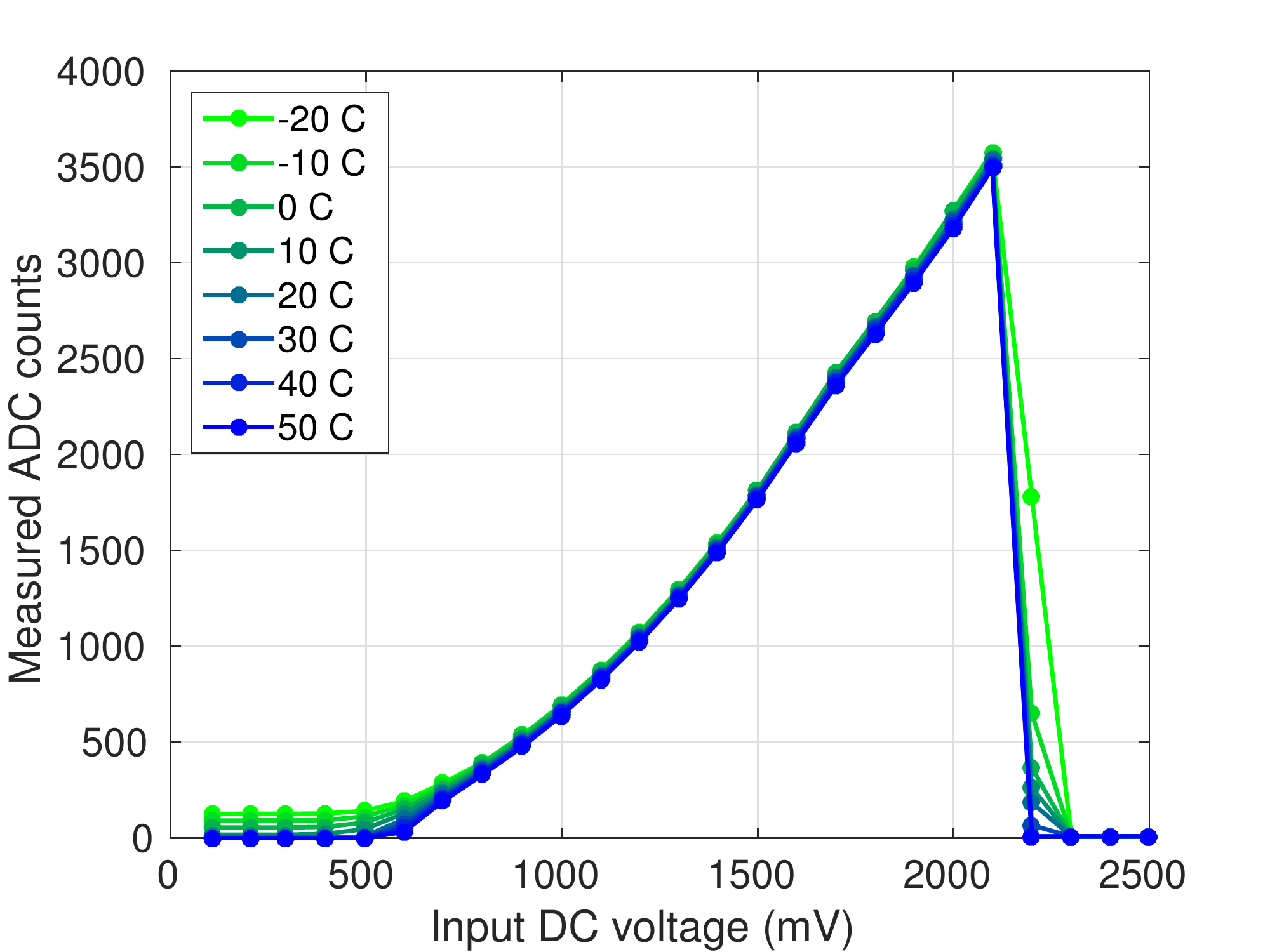}
\caption{Performance of control loop designed to stabilize transfer function against temperature variation.  The first panel shows the transfer function at various temperatures with feedback disabled.  The second panel shows the same curves with feedback enabled.}
\label{temperature}
\end{figure}

The Wilkinson counter frequency is selected by the \tp{Vdly} control voltage.  This circuitry has some temperature dependence, resulting in temperature dependence of the transfer function slope.  However, the Wilkinson clock frequency can be monitored and used for a control loop that varies \tp{Vdly} (provided by an {\mod internal} DAC) in order to stabilize the Wilkinson frequency despite varying temperature.  The transfer {\rev function was} measured at various temperatures in a thermal chamber, both with and without this feedback loop enabled.  The results are shown in Figure~\ref{temperature}.  After enabling feedback, there is a small amount of residual temperature dependence, likely due to {\mod temperature sensitivity of other parts of the digitization circuit}.


Figure~\ref{transfer-function} shows an example TARGET~5 transfer
function at room temperature for a typical configuration.  In general
the transfer function depends on both the configuration and on the
channel and cell position in the {\mod SCA}.  {\rev The integral nonlinearity of the transfer function shown in Figure~\ref{transfer-function} (with respect to the best linear fit, which has a slope of 2.5 counts/mV) is 212~ADC counts.  A calibration procedure is used to remove effects of this nonlinearity from data.   {Polynomial parameterization of the transfer
function (up through fourth order)} was evaluated and found to provide insufficient precision.  A
lookup table was found to be a more precise solution and {\rev is} implemented
in offline analysis software.  We found that 25 data points (spanning 0.1 to 2.5 V) are sufficient to specify its shape to a precision better than the DC noise.  Variation from cell to cell is included in the calibration procedure but is small, as indicated by the $\pm$1~$\sigma$ curves in Figure~\ref{transfer-function}.


{\rev The transfer function shown in Figure~\ref{transfer-function} is the result of optimizing for effective dynamic range.  The input range is 1.2~V (spanning {\mod 0.8 to 2.0}~V).  This input range spans 315 to 3243 ADC counts, corresponding to 2928 counts (11.5 bits) of resolution. The DC noise, after calibration of the transfer function, was measured (averaged over input DC voltage from 0.8 to 2.0 V) to be 0.6 mV (1.4 ADC counts, or 0.5 least-significant bits).  The effective dynamic range is therefore $11.5-0.5$, {\mod or, equivalently, $ \log_2 (1200/0.6)= 11.0$ bits}.  This compares favorably with TARGET 1, which has an effective dynamic range of 9.1 bits~\cite{2012APh....36..156B}.}

\begin{figure}
\includegraphics[width=0.5\textwidth]{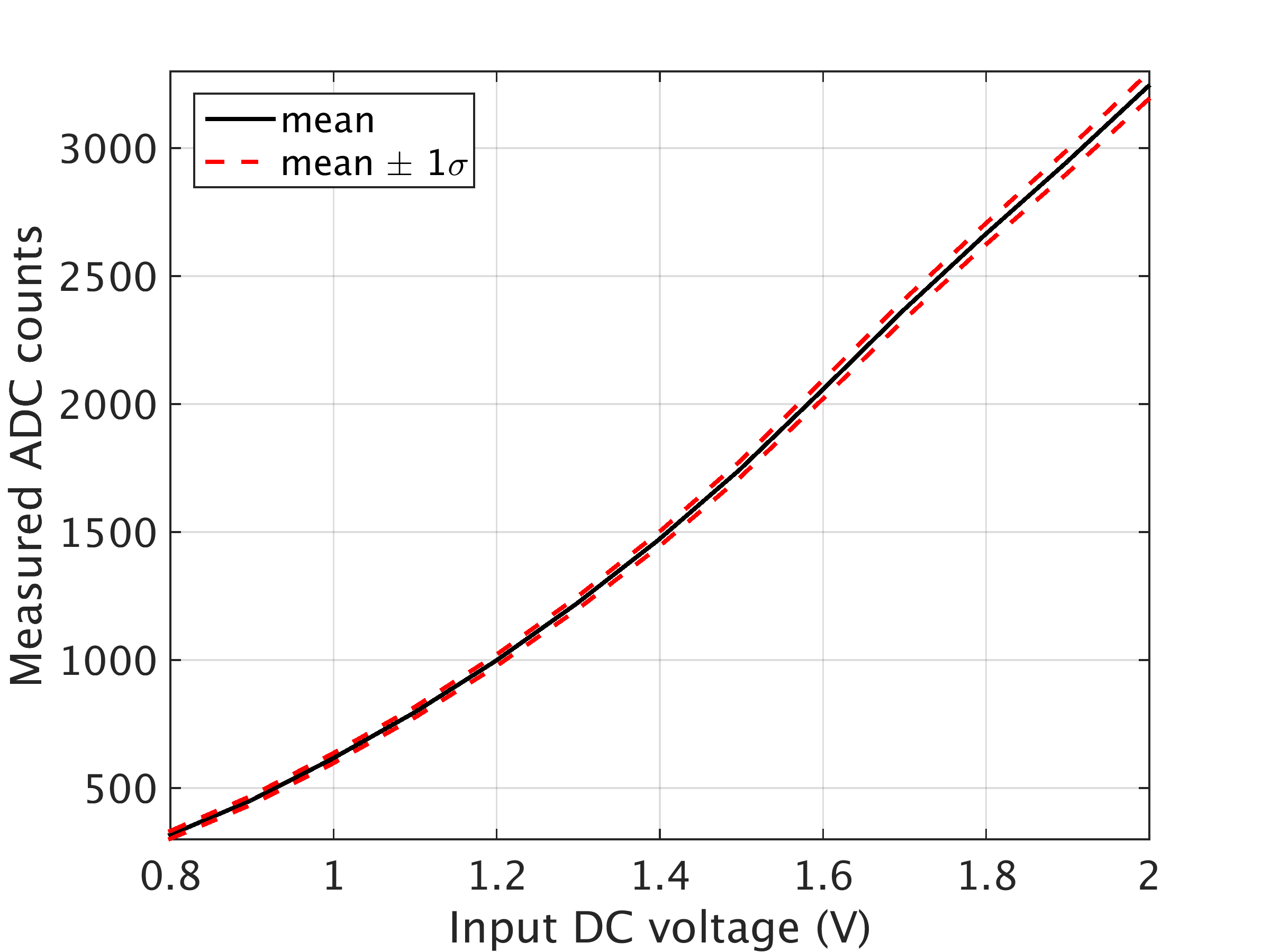}
\caption{\rev TARGET~5 transfer function for a single channel with 64-sample readout.  The 1~$\sigma$ variation among all 16,384 storage cells is shown.  Two different transfer functions are included for each cell, one when the cell occurs in the first block of the waveform and one when the cell occurs in the second block of the waveform, to include possible sequence dependence.}
\label{transfer-function}
\end{figure}

\subsubsection{ASIC response to sinusoids}\label{par-sinusoids}

The DC transfer functions described previously are used to convert measured ADC counts to instantaneous input voltage.  {\rev The ASIC response was also evaluated using sinusoidal signals from a function generator calibrated with the transfer functions determined from DC input.}  By scanning the input AC amplitude and comparing to the measured amplitude, the TARGET {\rev ``AC transfer function''} is determined (Figure~\ref{acSaturation}).  In TARGET~1, high frequency signals were found to exhibit an ``AC saturation'' effect at large amplitude~\cite{2012APh....36..156B}.  Note that this is not an effect due to finite bandwidth, which would cause the AC transfer function to have a slope less than unity but independent of input amplitude.  This effect was understood using simulations of TARGET~1 as due to an insufficient slew rate of the input buffer amplifiers and fixed in the design of subsequent revisions.  As shown in Figure~\ref{acSaturation}, AC saturation in TARGET~5 is negligible.

\begin{figure}
\includegraphics[width=0.5\textwidth]{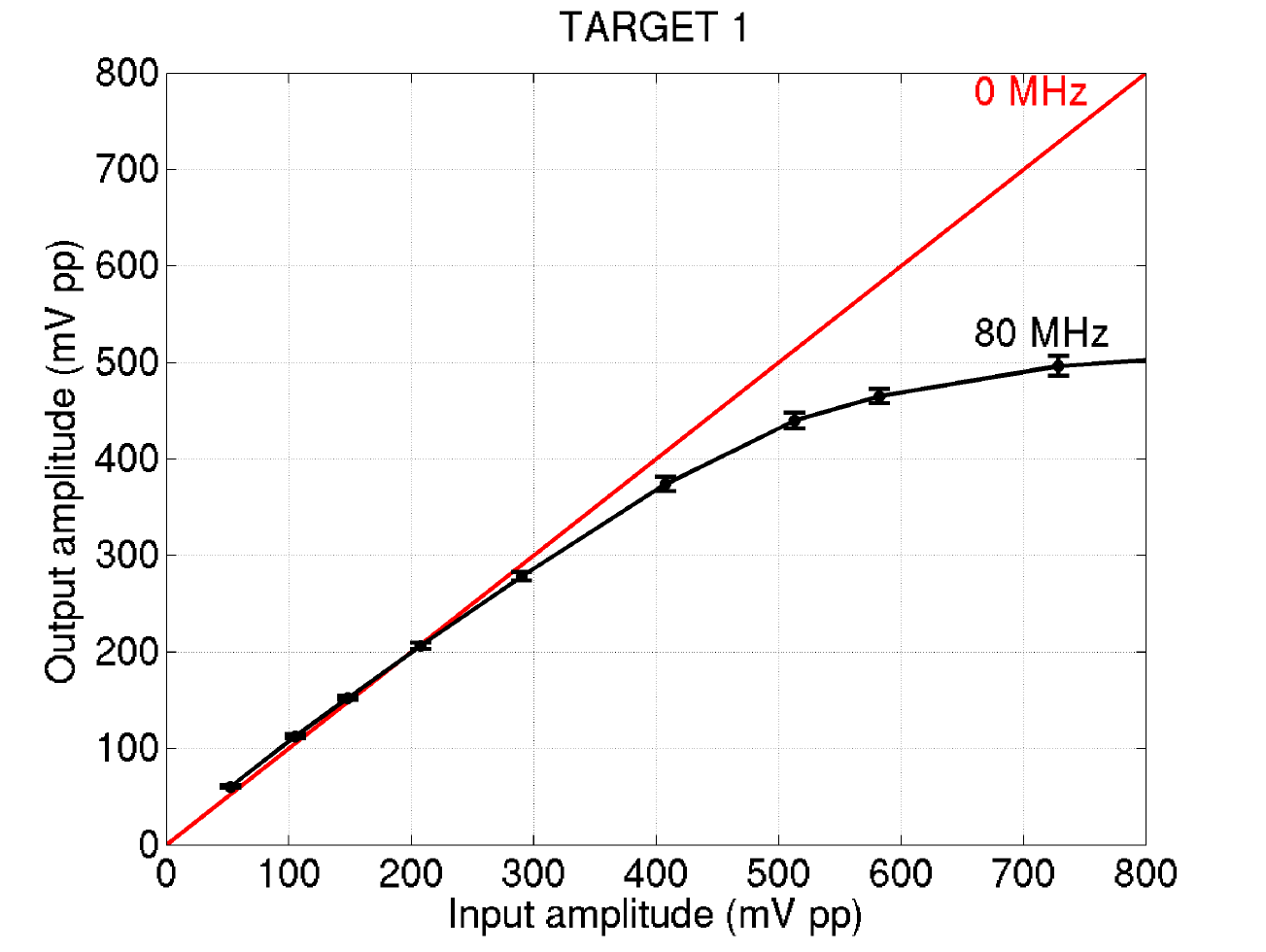}
\includegraphics[width=0.5\textwidth]{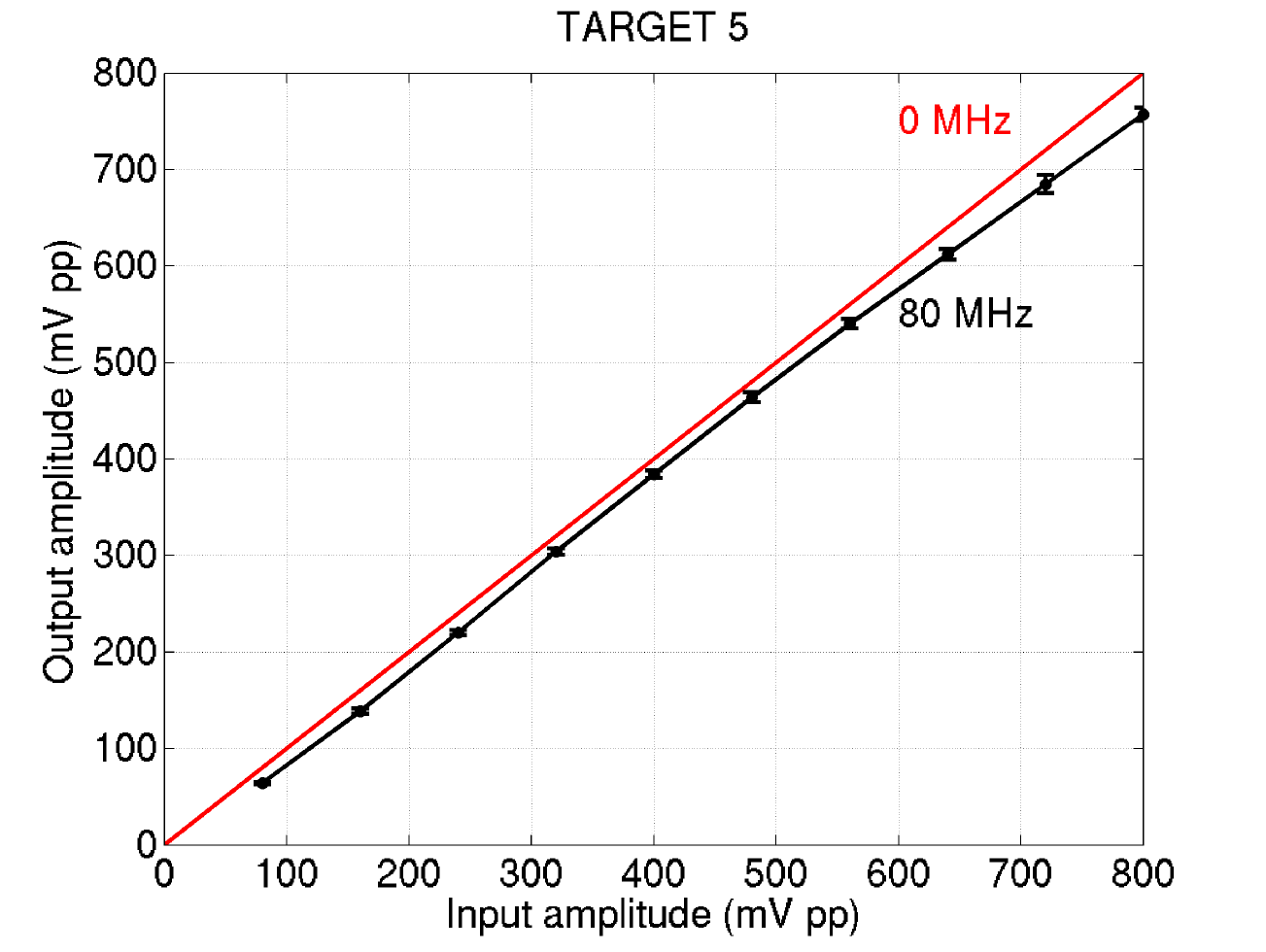}
\caption{{\rev Measured sinusoid amplitude, as a function of input amplitude,} for TARGET~1 and TARGET~5.  The AC saturation effect observed in TARGET~1 is negligible in TARGET~5.}\label{acSaturation}
\end{figure}



{\rev The effective noise measured with sinusoidal signals, i.e., the deviation of waveform samples from the fitted sinusoid, is larger than the noise measured with DC signals, as shown in Figure~\ref{acNoise}. }  This effective noise is caused by slightly suboptimal timing performance of the sampling, with a potential additional contribution due to settling of the signals from the sampling to the storage array and hysteresis in the sampling array. Therefore, it is not additive noise independent of the signal but it depends on the signal frequency and amplitude. 

For fixed signal frequency, this noise increases linearly with the signal amplitude, as shown in Figure~\ref{acNoise}.  The noise is 80\% at 5 mV signal amplitude and decreases to 5\% at 500 mV. {\rev The relationship between signal and noise is modeled well by a linear fit at each frequency, as indicated by the best-fit curves in Figure~\ref{acNoise}. The linear dependence supports the hypothesis that the main source of this noise is slightly suboptimal timing performance.  The effective noise measured in this configuration with small signal amplitude is $\sim$4~mV, independent of signal frequency.  This is larger than the 0.6~mV noise measured with DC signals and is most likely due to noise injected by the function generator.\footnote{This was confirmed with tests using a different model function generator.}}


{\rev The effective AC noise performance is acceptable} for measurements of integrated charge from photodetectors, an application for which photon counting is dominated by Poisson noise at the low amplitude range (limiting resolution to ${\sim}100$\%) and resolution of 10\% is typically sufficient for many-photoelectron signals.
\begin{figure}
\includegraphics[width=0.5\textwidth]{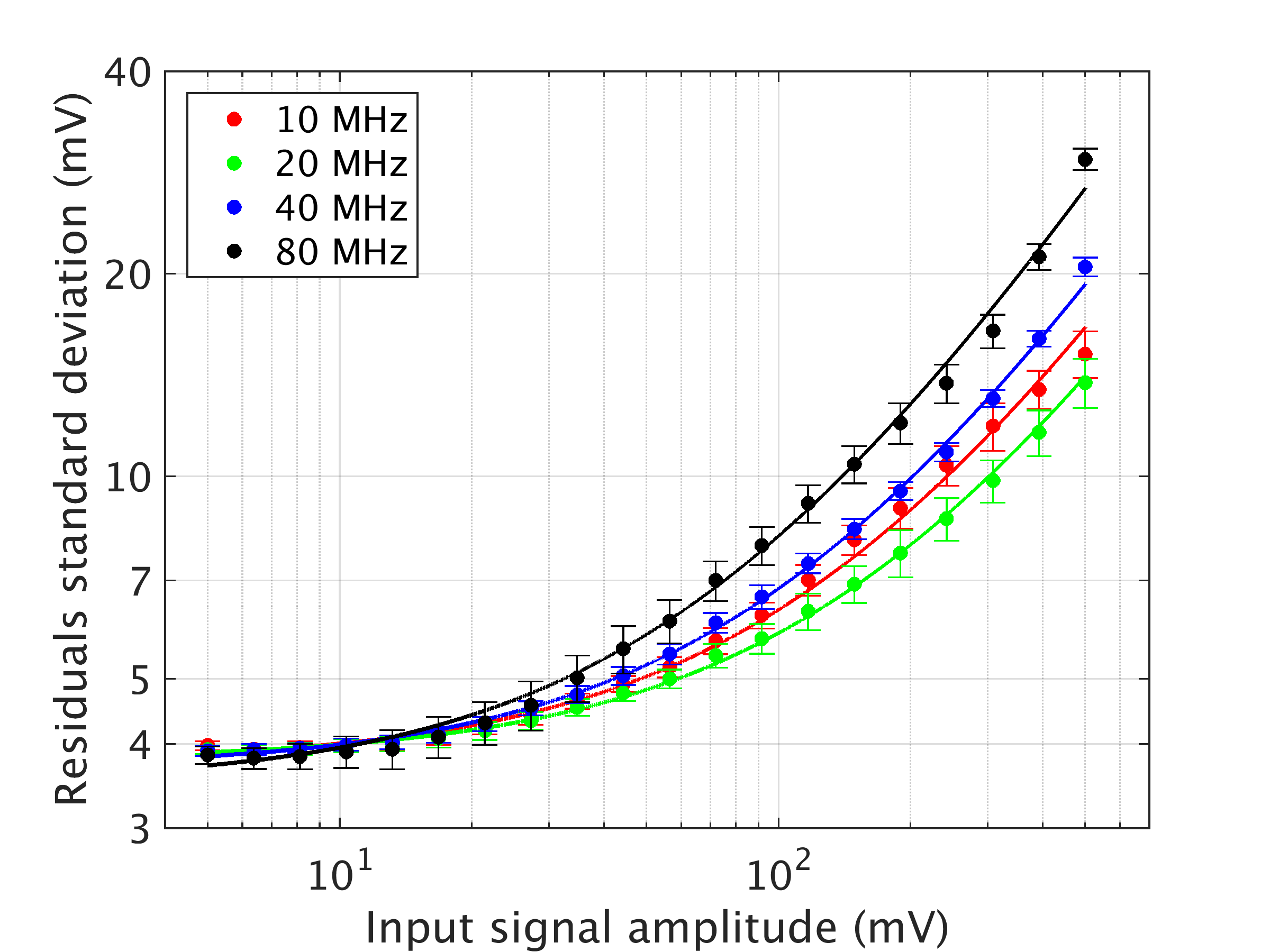}
\caption{\rev Effective AC noise measured with sinusoidal signals of various frequencies.  The input sinusoid amplitude was scanned from 10 mVpp (5 mV amplitude) to 1 Vpp.  The effective AC noise was measured by fitting sinusoids and then {\rev calculating} the standard deviation of the fit residuals. Curves show a linear fit to the effective noise as a function of input amplitude.}\label{acNoise}
\end{figure}

\subsubsection{Bandwidth}\label{par-bandwidth}

The bandwidth of the TARGET~5 ASIC was measured by comparing the amplitude of input sinusoids to the amplitude of waveforms recorded by the ASIC. To generate a range of sinusoidal input from ${\sim}50$~MHz to $>800$~MHz, we used four voltage-controlled oscillators (VCOs) whose variable frequencies and amplitudes were calibrated with a fast digital oscilloscope (Tektronix MDO4104-6, $1$~GHz 3-dB bandwidth). The amplitudes of recorded waveforms were estimated by calculating the standard deviation of thousands of voltage samples.

Figure~\ref{fig:bandwidth} shows the measured attenuation of TARGET 5 as a function of input frequency. In this measurement, three different attenuators ($3$, $6$, and $9$~dB) were used to repeat the same measurement with different input amplitudes, showing that different combinations of attenuators and VCOs are consistent with one another within ${\sim}0.2$~dB. The 3-dB bandwidth of TARGET~5 is approximately 500~MHz.

\begin{figure}
\includegraphics[width=0.5\textwidth]{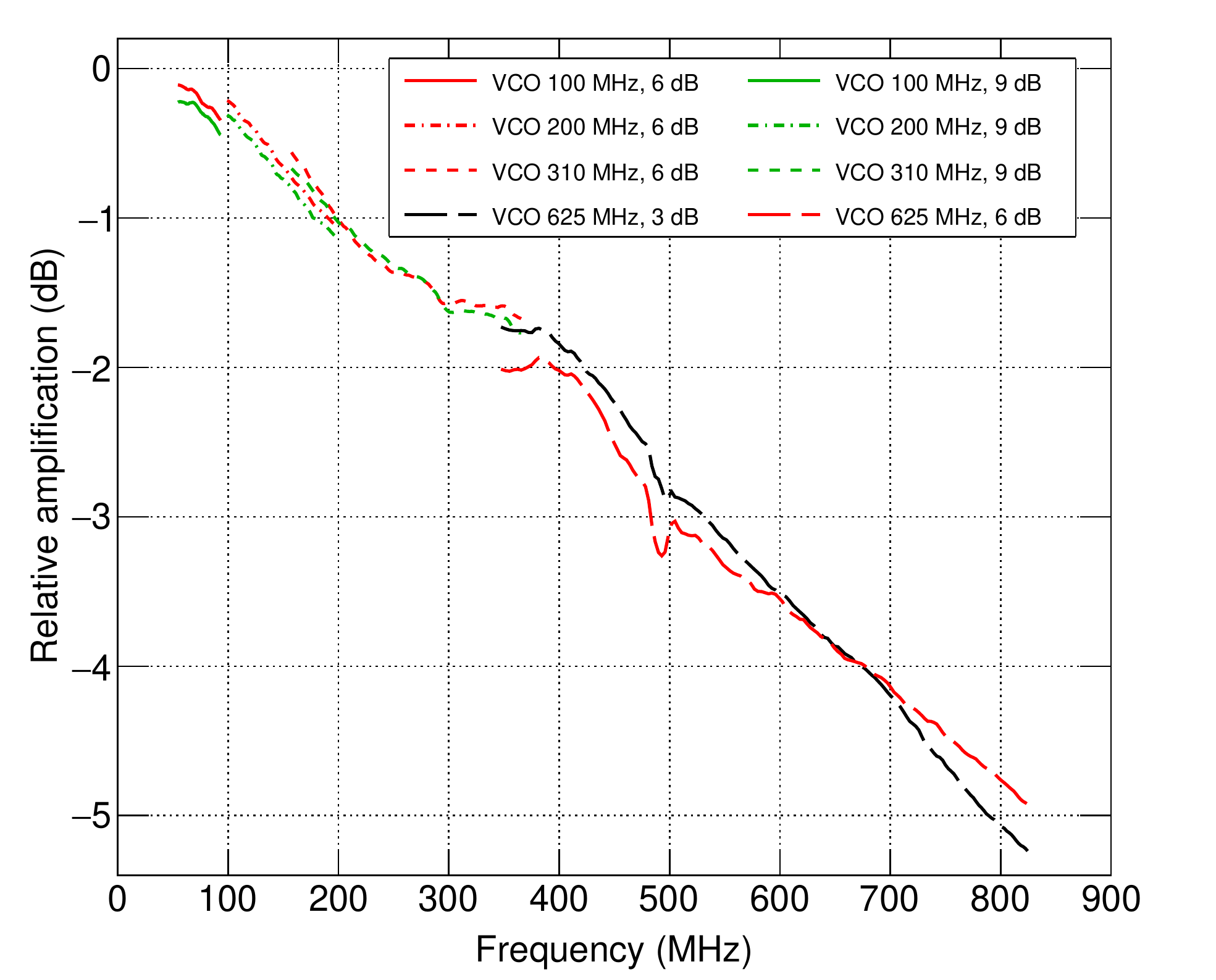}
\caption{The relative signal amplification of TARGET 5 as a function of input frequency. Different line styles and colors represent eight measurement configurations (four VCOs and three attenuators). The data points were smoothed by averaging over $\pm8$~MHz. The drop at ${\sim}490$~MHz is an artificial effect due to the measurement algorithm. Labels ``VCO 100 MHz'', ``200 MHz'', ``310 MHz'', and ``625 MHz'' in the legend denote VCOs with different frequency ranges: $40$--$100$ MHz (Mini-Circuits ZX95-100), $90$--$220$ MHz (ZX95-200), $140$--$380$ MHz (ZX95-310A), and $340$--$840$ MHz (ZX95-625A).}
\label{fig:bandwidth}
\end{figure}

{\rev

\subsubsection{Crosstalk}\label{par-xtalk}

Due to the layout of the ASIC, AC signals induce a small amount of crosstalk on nearby channels. The measurements described in~Sections \ref{par-sinusoids} and~\ref{par-bandwidth} provide two complementary estimates of the crosstalk. In both cases, continuous wave sinusoids of various frequencies were injected to individual channels and we read out all channels.  When using the function generator (\ref{par-sinusoids}) to estimate the crosstalk we used a large input amplitude (0.8~V peak to peak),  and we performed sinusoid fits to all channels, fixing the frequency to the known value. In the case of the VCOs (\ref{par-bandwidth}), which extend the measurement to higher frequencies (hence, smaller measured amplitudes), the crosstalk was estimated from the standard deviation of  thousands of voltage samples.

Figure~\ref{crosstalk} shows a summary of the results from the two crosstalk estimates. The two measurements show good quantitative agreement in the frequency range where they overlap between 30 and 160~MHz, where the crosstalk ratio is at most 0.5\%. The crosstalk was measured to be largest on nearest neighbors, followed by  next-to-nearest neighbors as expected from the layout of the ASIC. The crosstalk ratio shows a marked frequency dependency and below 500 MHz (3dB bandwidth of the ASIC and Nyquist frequency for typical 1 GSa/sec operation) it is at most 1.3\%.

\begin{figure}
\includegraphics[width=0.5\textwidth]{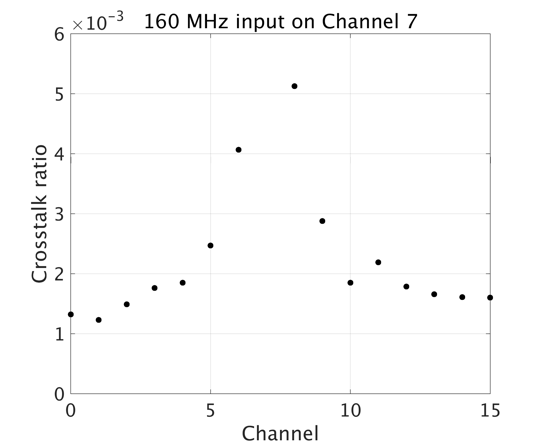}
\includegraphics[width=0.5\textwidth]{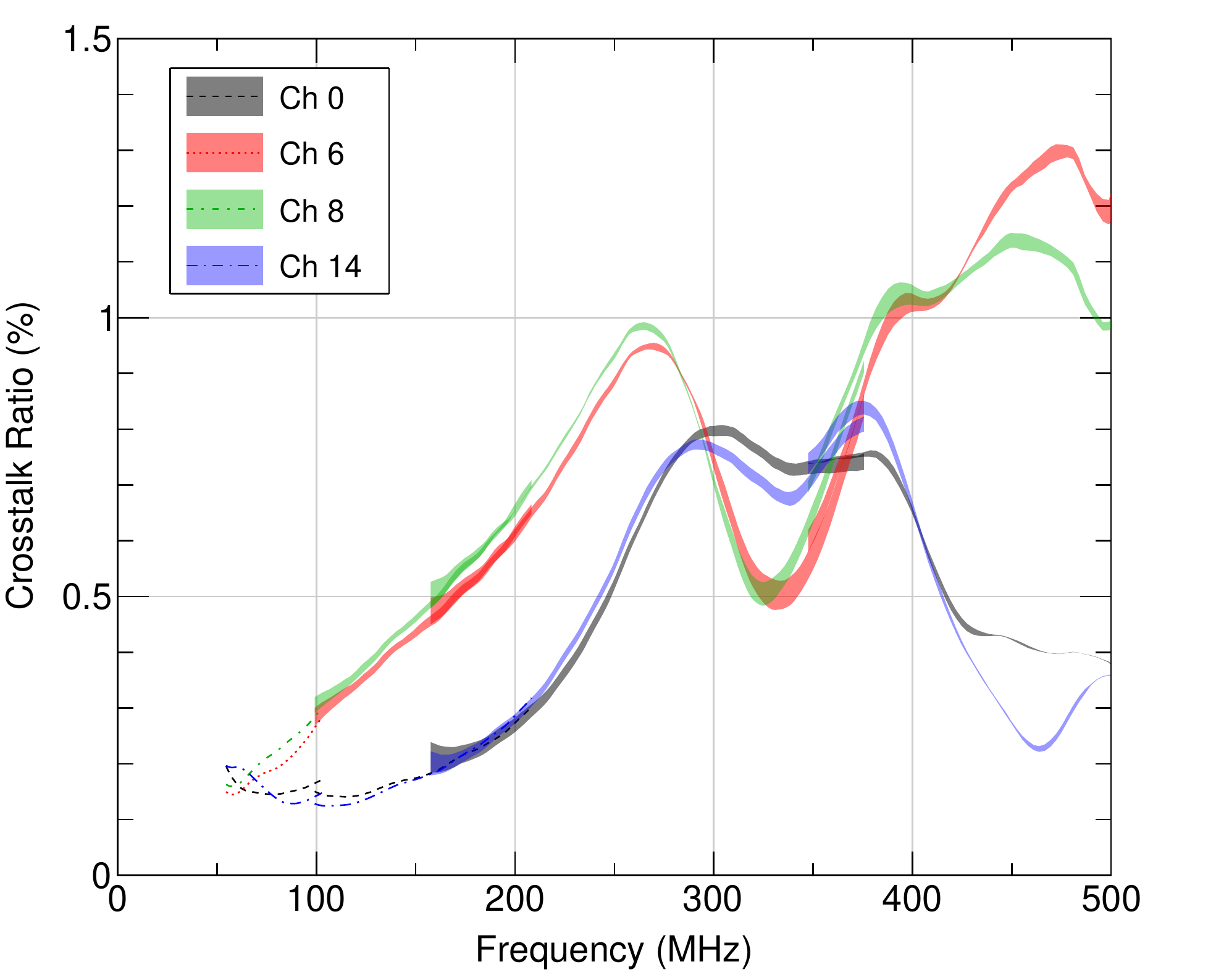}
\caption{\rev Crosstalk ratio (crosstalk amplitude divided by the input signal amplitude) for sinusoidal signals input to Channel 7. Top: crosstalk ratio for input frequency of 160 MHz as a function of channel number.  Note that pure DC noise corresponds to an expected crosstalk ratio $\sim$1$\times10^{-3}$. The measurement is described in Section~\ref{par-sinusoids}. Bottom: crosstalk ratio as a function of sinusoid frequency for a few representative channels. The bands encompass systematic differences from using different VCOs and attenuators. The measurement is described in Section~\ref{par-bandwidth}.}
\label{crosstalk}
\end{figure}

}

\subsubsection{ASIC response to pulses}

In addition to quantifying the performance of TARGET~5 with sinusoidal signals, we studied the performance with short electrical pulses (from a function generator) similar to those produced by a photodetector after shaping through a preamplifier as planned in CHEC-M \cite{2013arXiv1307.2807D,2015arXiv150901480D} and the SCT~\cite{SCTICRC2015}.  This is especially useful for quantifying the impact of the effective noise in terms of photoelectron charge resolution.  Each pulse had an 8~ns {\rev full width at half maximum}, 5~ns rise time, and 5~ns fall time.  The amplitude was varied to {\mod emulate} variation in the number of photoelectrons (p.e.).

A charge reconstruction algorithm was applied as follows {\rev to digitized waveforms recorded using a sampling frequency of 1~Gsa/s}.  The 17~samples centered on the peak sample are determined and designated as on-pulse.  Eight samples before these on-pulse samples are avoided because they lie in the transition region between on-pulse and off-pulse.  Furthermore, all samples after the on-pulse samples are avoided because they could include undershoot.  The remainder of the samples, well before the pulse, are designated off-pulse.  An example waveform with this algorithm applied is shown in Figure~\ref{pulse}.  The mean of the off-pulse samples is calculated and used as a baseline estimate.  The mean of the on-pulse samples is then determined and the baseline is subtracted.  The resulting quantity provides an estimate of the baseline-subtracted charge integral in the pulse and increases linearly with the pulse input amplitude.  This quantity is calibrated to photoelectrons assuming a preamplifier gain of 4~mV per photoelectron.  The resulting measured charge as a function of simulated input charge is shown in Figure~\ref{charge-linearity}.

The charge resolution is quantified for each input pulse amplitude by the standard deviation of the reconstructed charge.  The relative charge resolution is shown in Figure~\ref{charge-resolution}.  This charge resolution is dominated by the {\rev effective AC noise described in Section \ref{par-sinusoids}. In particular, the increase of effective noise at large amplitudes (Figure~\ref{acNoise}) causes the charge resolution to plateau for large charge values.} In a full system consisting of photodetector plus readout electronics, the charge resolution will be worse than this due to other contributions including photodetector noise, crosstalk, after-pulses, gain uncertainty, and Poisson noise.

\begin{figure}
\includegraphics[width=0.5\textwidth]{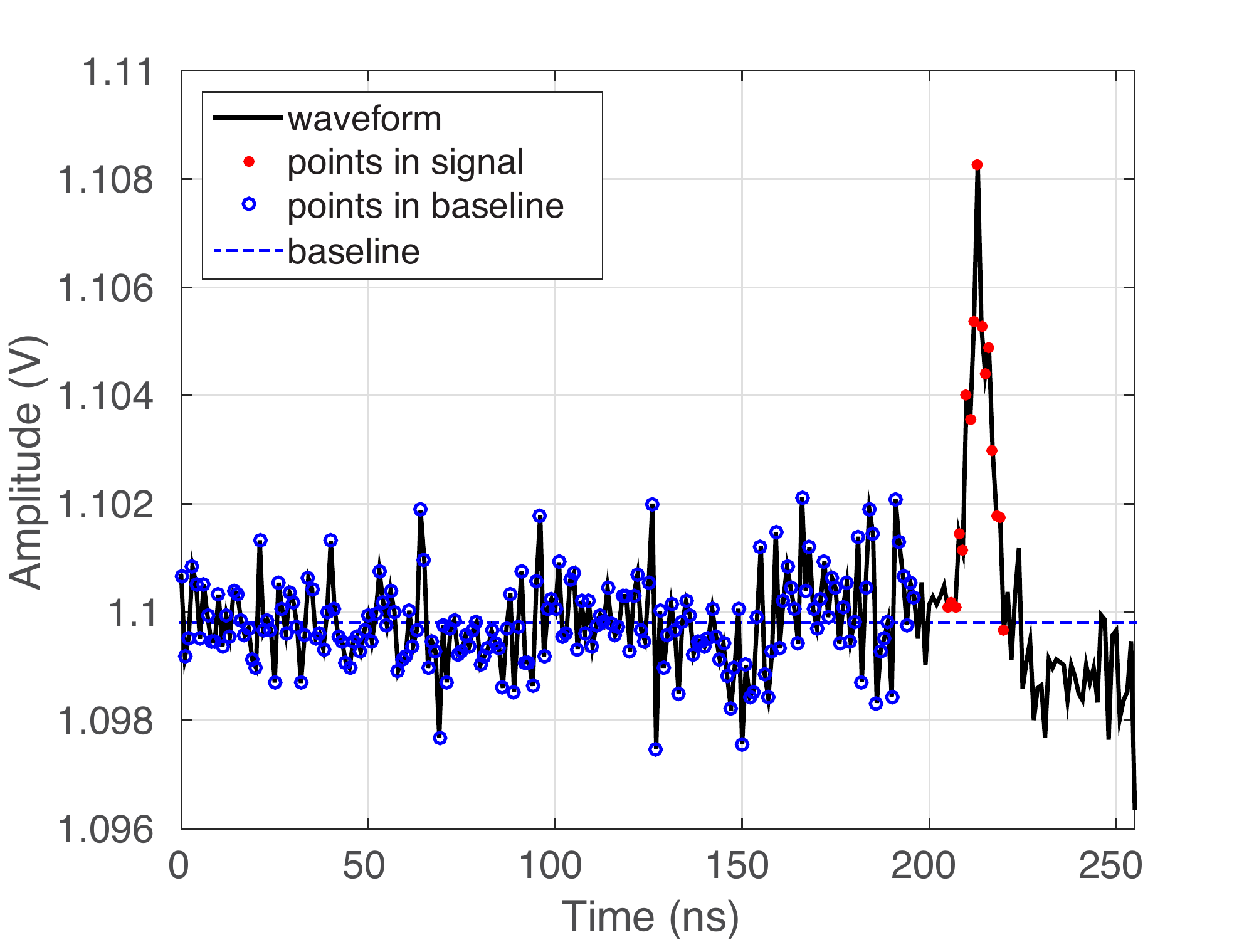}
\caption{Example electrical pulse recorded with the TARGET~5 evaluation board.  The input amplitude was 10 mV, corresponding to 2.5 photoelectrons for an example preamplifier gain of 4~mV per photoelectron.}\label{pulse}
\end{figure}

\begin{figure}
\begin{centering}
\includegraphics[width=0.5\textwidth]{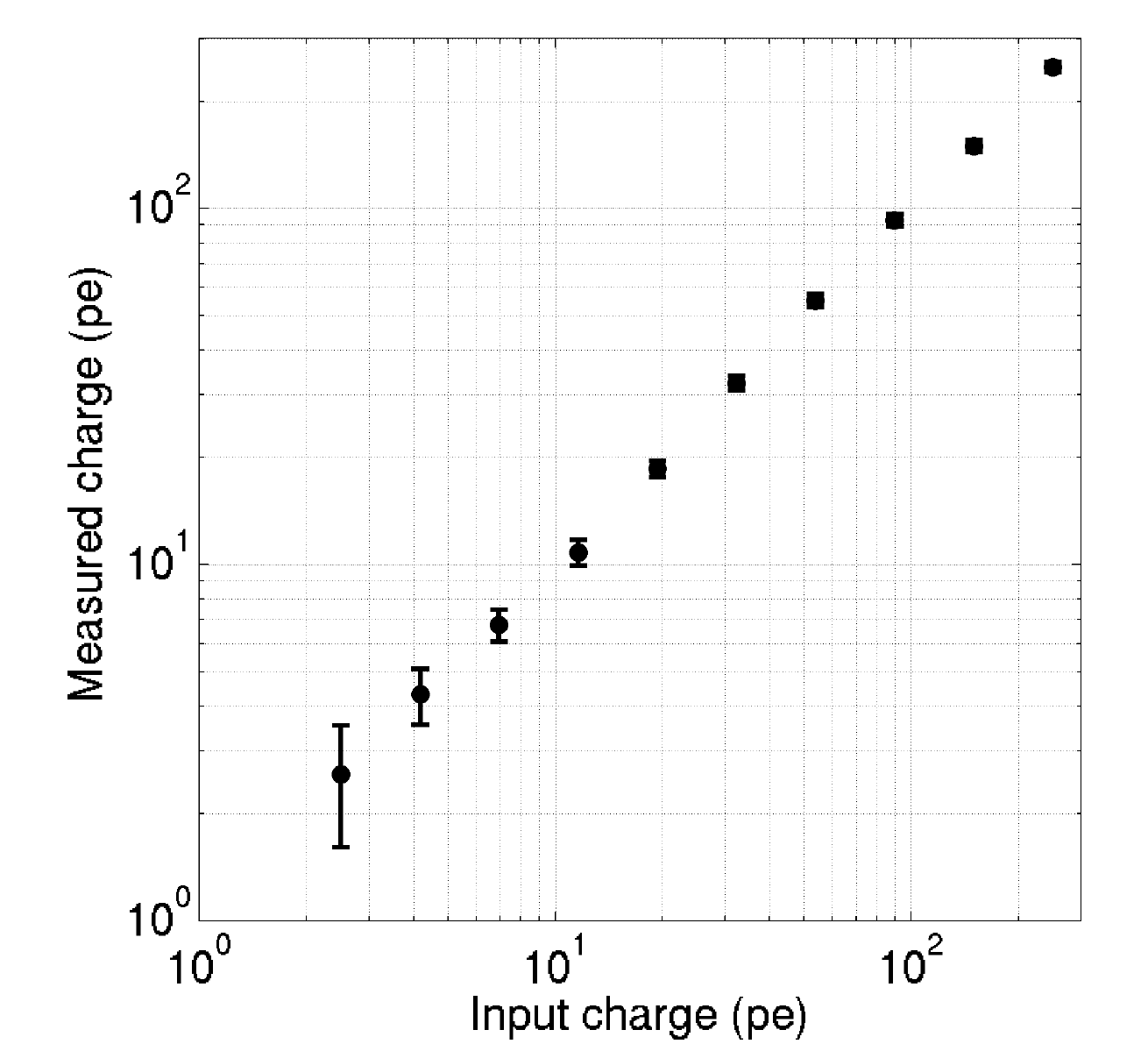}
\caption{Linearity of the TARGET~5 pulse response.  Photoelectron pulses were simulated using electrical pulses with an assumed gain of 4~mV per photoelectron.  The measured charge was determined by integrating sample amplitudes.  Error bars indicate the $\pm1~\sigma$ variation in reconstructed charge.}
\label{charge-linearity}
\end{centering}
\end{figure}

\begin{figure}
\includegraphics[width=0.5\textwidth]{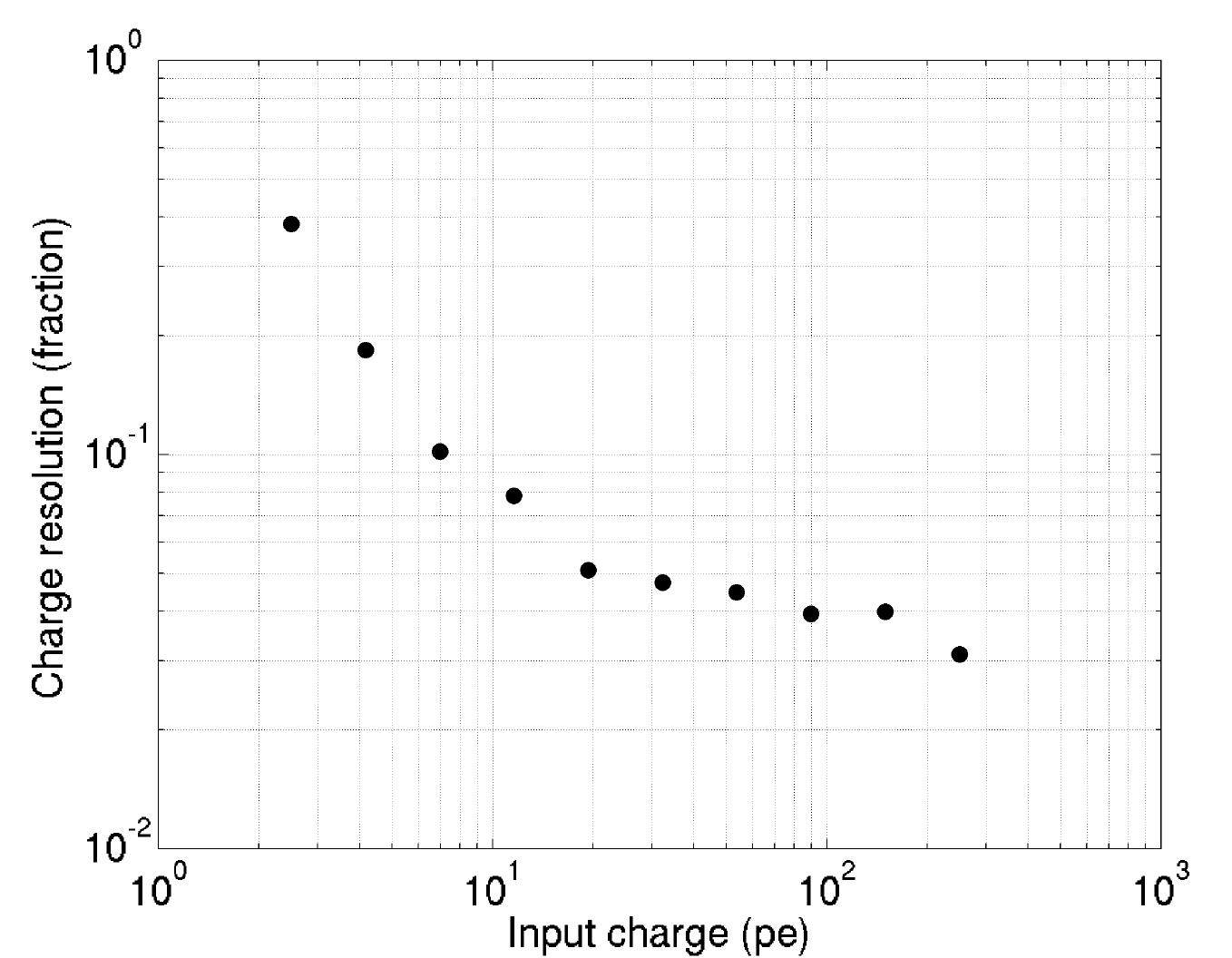}
\caption{TARGET~5 charge resolution.  The charge resolution was determined at each input amplitude by dividing the standard deviation of the reconstructed charge by the mean value.}
\label{charge-resolution}
\end{figure}

\subsection{Trigger path}\label{sec:trigpath}

\subsubsection{Tuning of the trigger performance}
The trigger performance is determined in each trigger group mainly by two parameters, 
\tp{PMTref4}, which sets the reference voltage for the summing amplifier that performs the 
analog sum of the signal from four adjacent channels, and \tp{Thresh}, which sets the reference 
voltage for the comparator. Both parameters set the voltages with respect to the {\rev corresponding} supply voltages (also tunable) and are controlled by a 
12-bit DAC.

To characterize the trigger performance, we performed a scan over these 
parameters. For each setting we injected in one channel pulses of variable 
amplitude with frequency of 1 kHz, full width at half maximum of 8 ns, and 
{\modakira rise} time of 5 ns. For each pulse amplitude we used a counter implemented in the FPGA in order to 
count the number of trigger signals issued by the group to which the channel 
belongs over a time of 2.15~s. For simplicity, digitization and readout of 
waveforms was disabled during these {\rev scans of a large phase space of trigger configuration \footnote{\rev This prevents unstable trigger configurations from causing a large number of readout requests which breaks communication with the computer.}}.

We can estimate the trigger efficiency {\mod $\varepsilon = 
N_\mathrm{trig}/N_\mathrm{gen}$, where $N_\mathrm{trig}$ is the number of trigger signals issued by the ASIC as counted 
by the FPGA, and $N_\mathrm{gen}$} is the number of pulses generated by the function 
generator (i.e., the product of the pulse frequency and the counting time). 
The number of trigger signals generated follows a binomial distribution, where 
$\varepsilon$ corresponds to the probability of a pulse initiating a trigger 
signal.  Hence, the error on the efficiency is estimated\footnote{When 
$\varepsilon \sim 0$ or $\varepsilon \sim 1$ (the number of triggers issued 
{\modakira $N_\mathrm{trig}$} is {\modakira $\sim0$ or \modakira $\sim N_\mathrm{gen}$}) we approximately evaluate the trigger efficiency 
uncertainty by using the formulas {\modakira $\varepsilon = 
(N_\mathrm{trig}+1/3)/(N_\mathrm{gen}+2/3)$ and $\sigma_\varepsilon = \sqrt{\varepsilon 
(1-\varepsilon)/(N_\mathrm{gen}+2)}$ from \cite{2012JInst...7.8021C}.}} as 
{\modakira $\sigma_\varepsilon = \sqrt{\varepsilon (1-\varepsilon)/N_\mathrm{gen}}$}.

The trigger efficiency $\varepsilon$ as a function of pulse amplitude, $a$, for a 
particular configuration of the ASIC parameters is shown in Figure~\ref{trigger-Scurve-normal}. We fit to the efficiency points a function of the 
form
\begin{equation}
 S(a;\mu,\sigma)=\frac{1}{2}\left[1+\mathrm{erf}\left(\frac{a-\mu}{\sqrt{2}\sigma}\right)\right]
\end{equation}
where erf is the Gaussian error function, and $\mu$ and $\sigma$ are the fit parameters which represent the trigger threshold and noise, respectively.
\begin{figure}
\includegraphics[width=0.5\textwidth]{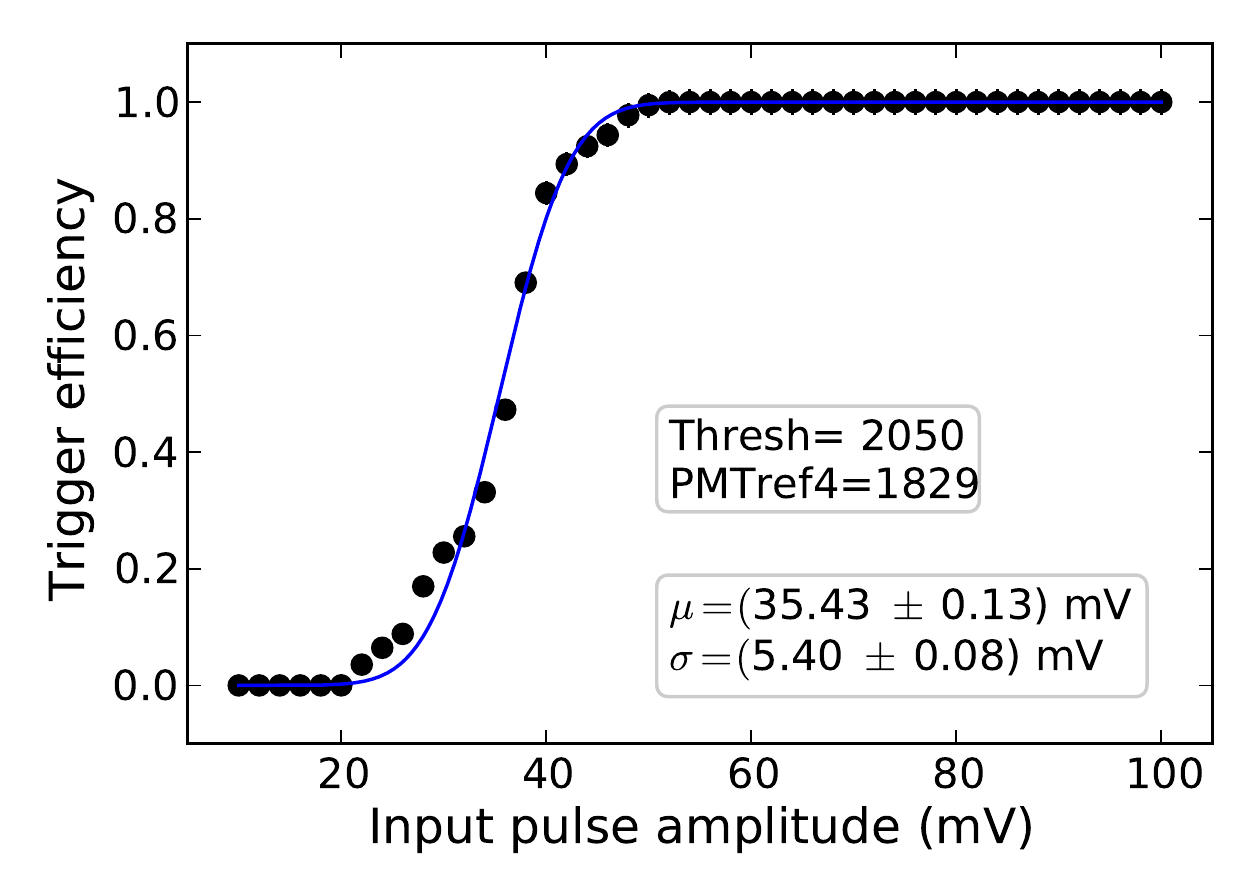}
\caption{The black points show the trigger efficiency measured as a function of 
input pulse amplitude during normal operations (with sampling enabled). The blue curve 
shows the best-fit $S$ function. The inset text gives the trigger configuration parameter values as well as
best-fit values of the trigger threshold $\mu$ and noise $\sigma$.}
\label{trigger-Scurve-normal}
\end{figure}

Figure~\ref{trigger-perf-normal} shows trigger threshold and noise from our scan 
of \tp{PMTref4} and \tp{Thresh}.
\begin{figure}
\includegraphics[width=0.5\textwidth]{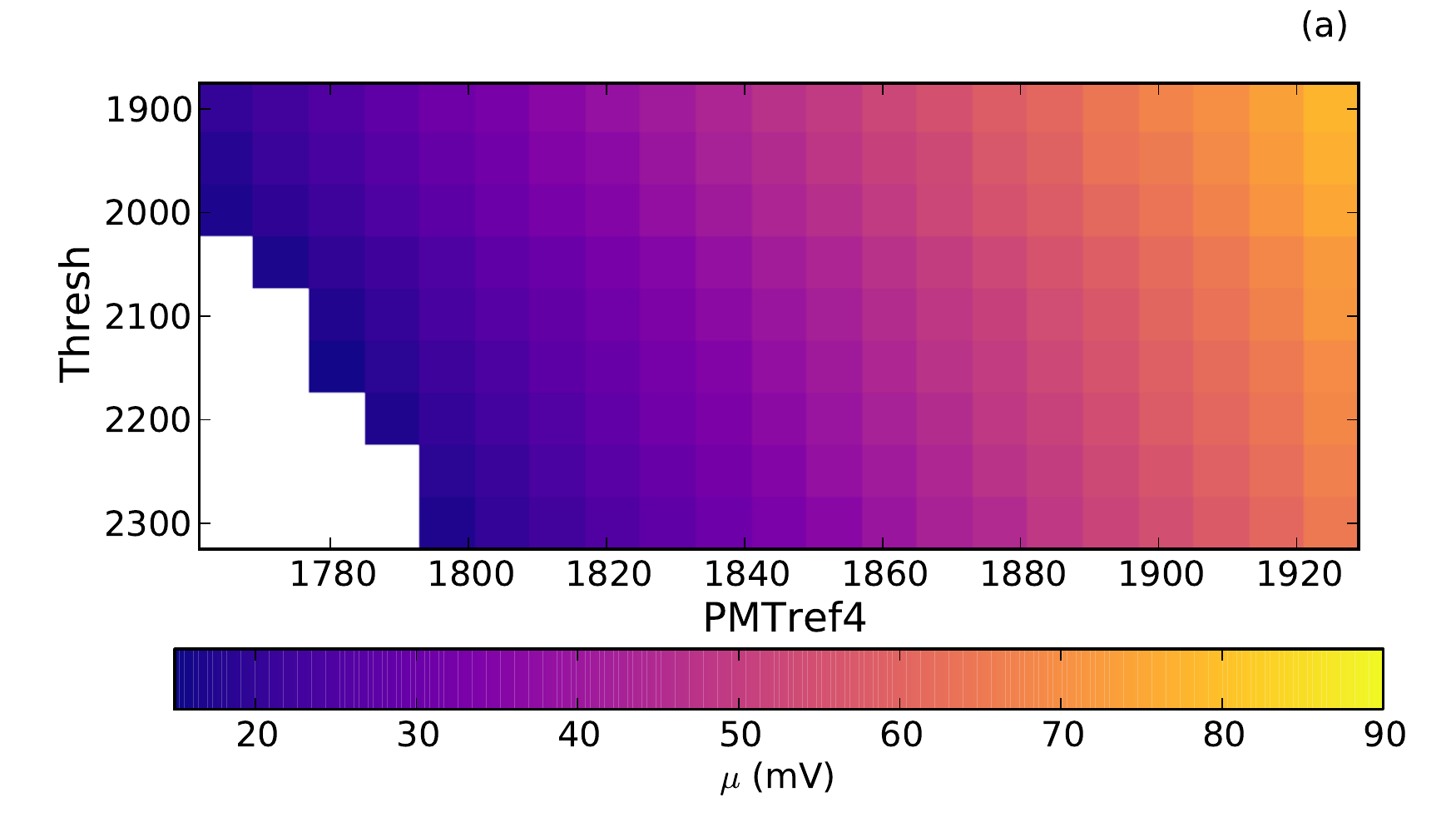} 
\includegraphics[width=0.5\textwidth]{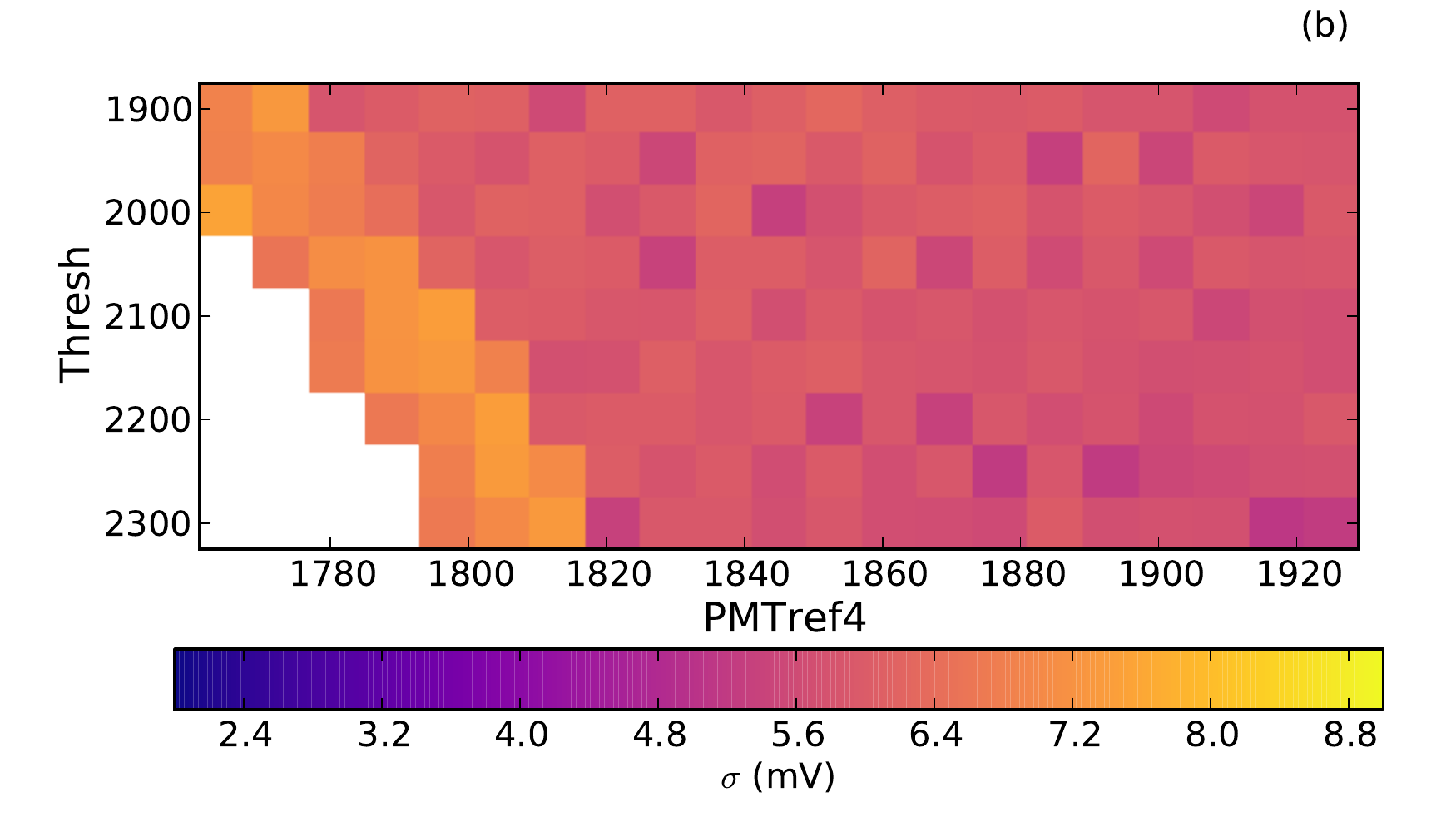}
\caption{Trigger threshold (a) and noise (b) as a function of \tp{PMTref4} and \tp{Thresh} (in DAC counts). White indicates a 
region of parameter space where the trigger does not function properly.}
\label{trigger-perf-normal}
\end{figure}
The minimum trigger threshold achievable in this mode (with analog sampling of the data path enabled) is ${\sim}20$~mV, 
with a trigger noise $\gtrsim 5$~mV, which correspond to 5~p.e. and 1.2~p.e., respectively, for a pre-amplifier with a gain of 4 mV per photoelectron. {\rev This performance does not meet the design goal of triggering on $\sim$2 photoelectrons with trigger noise below 1 photoelectron}.
The next paragraphs describe how we investigated the causes of this behavior.

\subsubsection{Trigger performance for operations in sync with sampling}

Since the input signal is going on one side to the trigger circuit, and, on the 
other side, to the sampling circuit that is the first stage of the data path, 
we wanted to study how the trigger performance depends on the sampling signals 
internal to the ASIC.

The input signal is sampled using as a reference a clock signal with a full 
period of 64~ns. For testing purposes, we extracted from the FPGA a signal 
synchronous with the sampling clock downscaled to a frequency of $\sim$119~Hz (generated every 512 full cycles through the 16,384 cells of the storage buffer). This signal was sent to the function generator as an external trigger, 
in order to generate pulses at fixed phase with respect to the sampling clock. 
We varied the delay between the external trigger to the function generator and 
the output pulse between 0 and 128~ns (2 full periods of the sampling clock), 
and evaluated for a given \tp{PMTref4}-\tp{Thresh} pair trigger threshold and noise as a 
function of this delay.

Figure~\ref{trigger-perf-sync} shows trigger threshold and noise from the delay 
scan.
\begin{figure}
\includegraphics[width=0.5\textwidth]{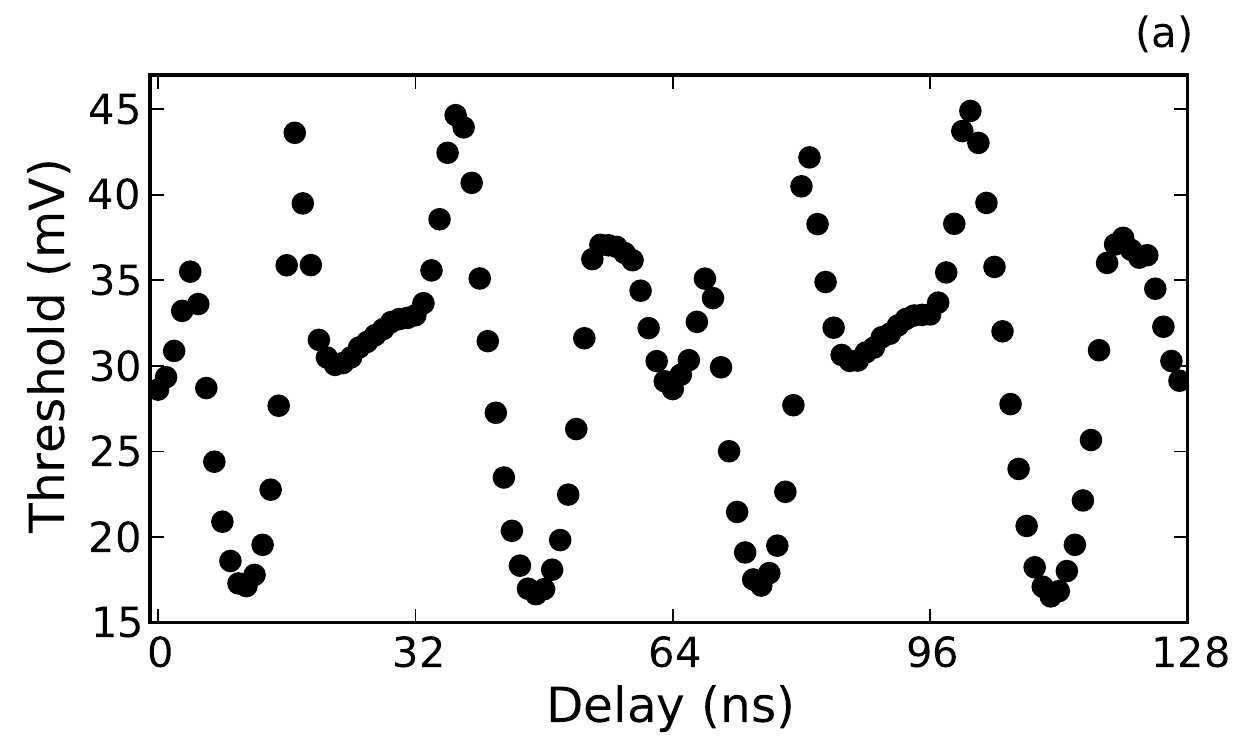} 
\includegraphics[width=0.5\textwidth]{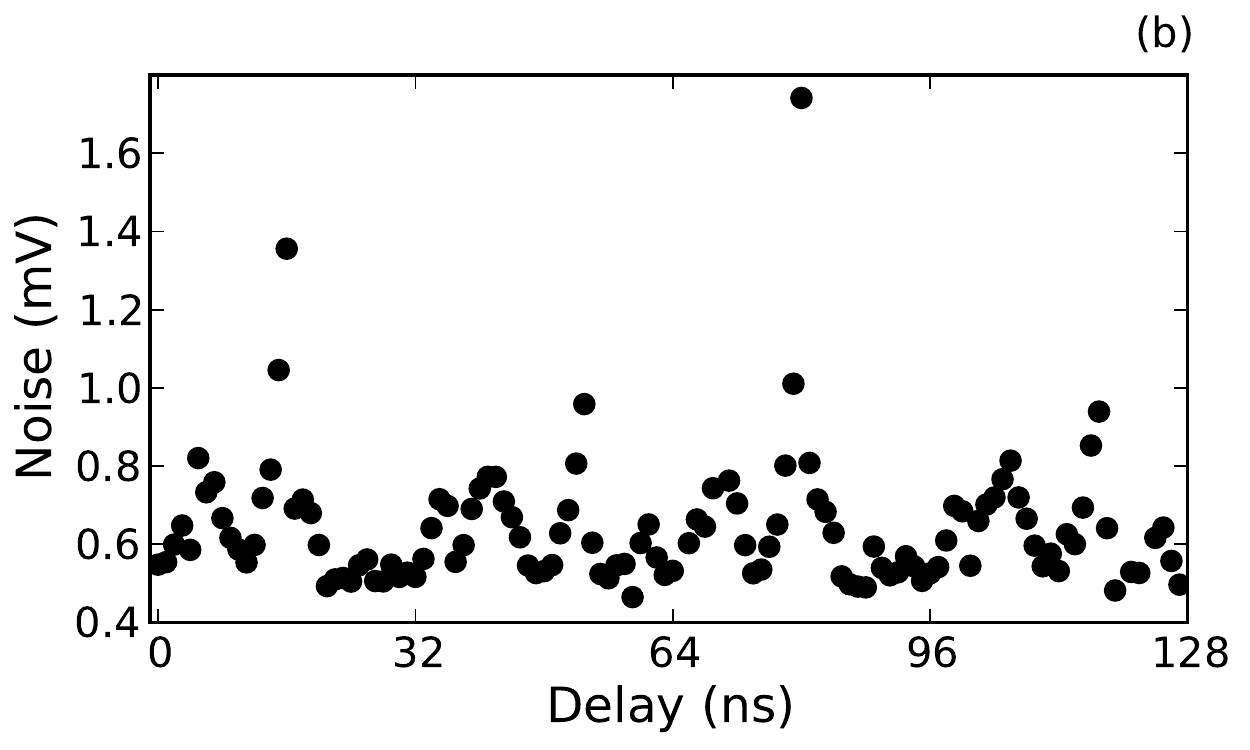}
\caption{Trigger threshold (a) and noise (b) as a function of 
delay between a signal synchronous with the 
sampling clock and the generation of the input 
pulse to the ASIC.}\label{trigger-perf-sync}
\end{figure}
The threshold exhibits a pattern with 64 ns period, clearly indicative of some 
interplay between sampling and triggering. The variations in the threshold 
values are as large as ${\sim}30$~mV. On the other hand, for operations in sync 
with the sampling clock the noise is typically $\lesssim 1$~mV, with larger 
values of a few mV in points where the threshold is rapidly changing. This 
demonstrates that the $\gtrsim 5$~mV noise in normal (asynchronous) operations{,\rev\ shown Figures~\ref{trigger-Scurve-normal} and \ref{trigger-perf-normal}(b),}
is mostly due to threshold variations for pulses that arrive at different 
sampling phases. 

Figure~\ref{figures/trigger-Scurve-sync.pdf} shows a typical efficiency curve 
for operations in sync with sampling. {\rev The transition is much sharper, 
and} in this case the $S$ function provides a much better fit to the data, 
which indicates that the deviations in normal mode are given by the 
superposition of many $S$ curves with different thresholds. 

\begin{figure}
\includegraphics[width=0.5\textwidth]{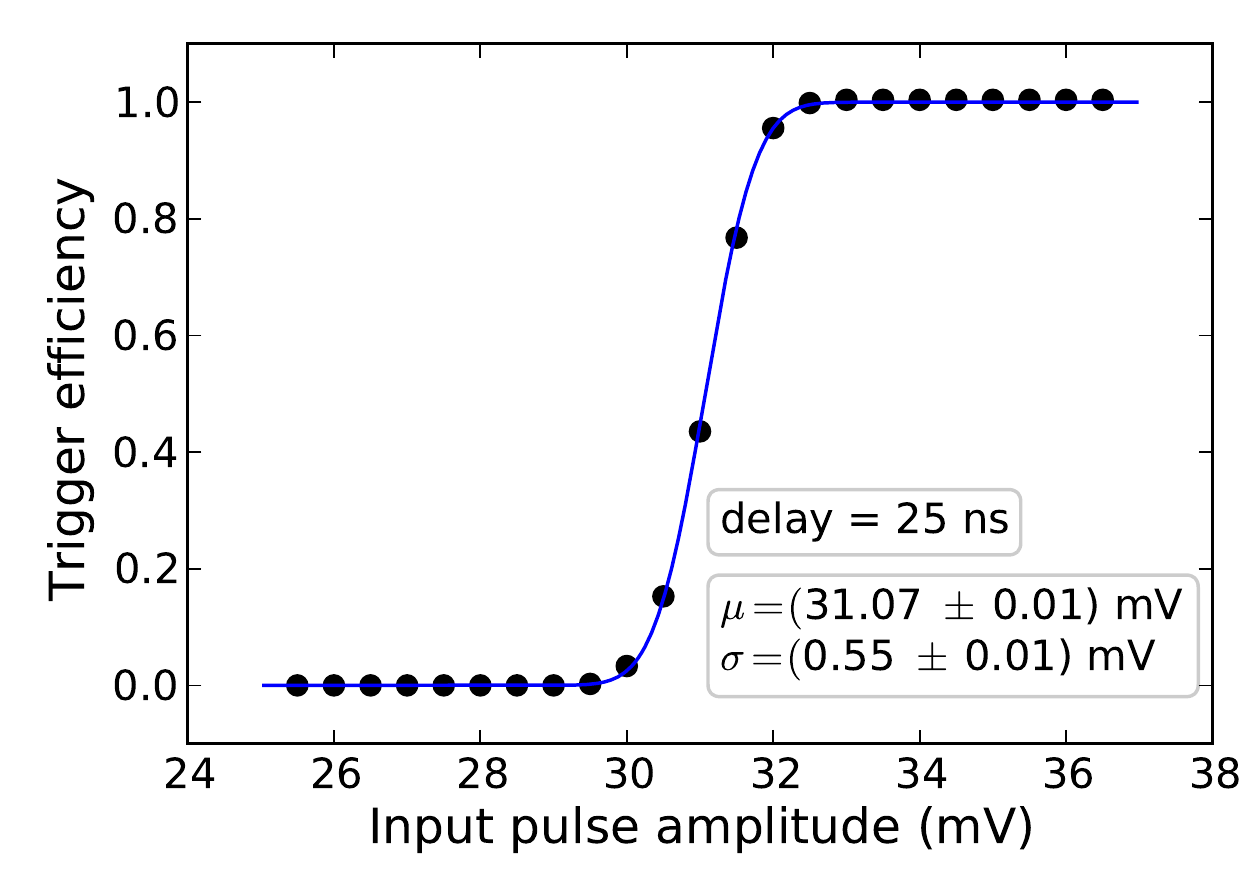}
\caption{Same as Fig.~\ref{trigger-Scurve-normal}, but the input pulse timing is synchronized with the signal sampling clock. The delay between the reference signal and input pulse is $25$~ns in this figure.}
\label{figures/trigger-Scurve-sync.pdf}
\end{figure}

\subsubsection{Trigger performance with sampling disabled}

Because the performance of the trigger system is strongly dependent on the 
analog sampling, we characterized the trigger performance with sampling disabled.
Figure~\ref{trigger-perf-samploff} shows trigger threshold and noise from our scan 
in \tp{PMTref4} and \tp{Thresh} when the sampling is disabled. 

\begin{figure}
\includegraphics[width=0.5\textwidth]{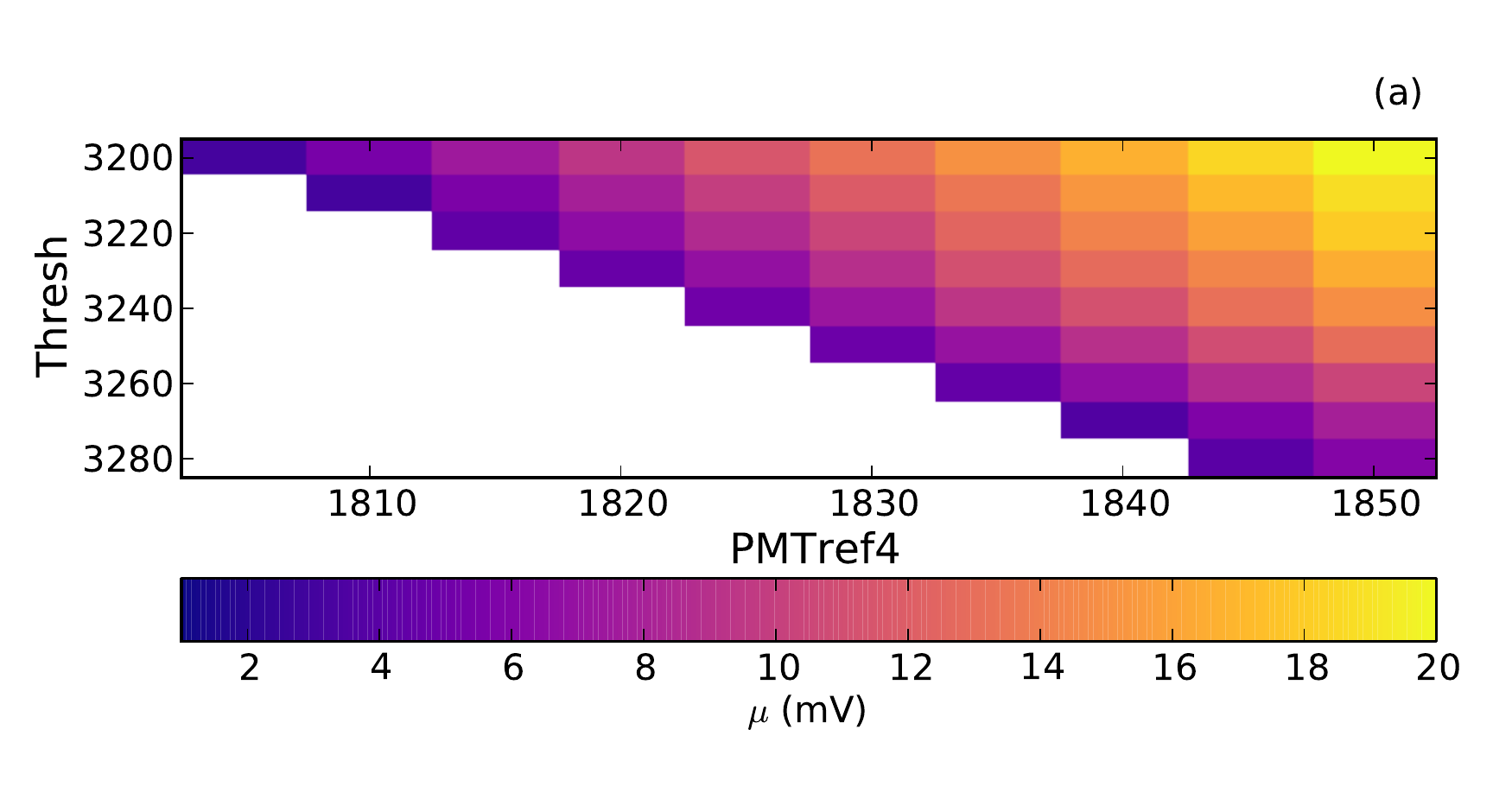} 
\includegraphics[width=0.5\textwidth]{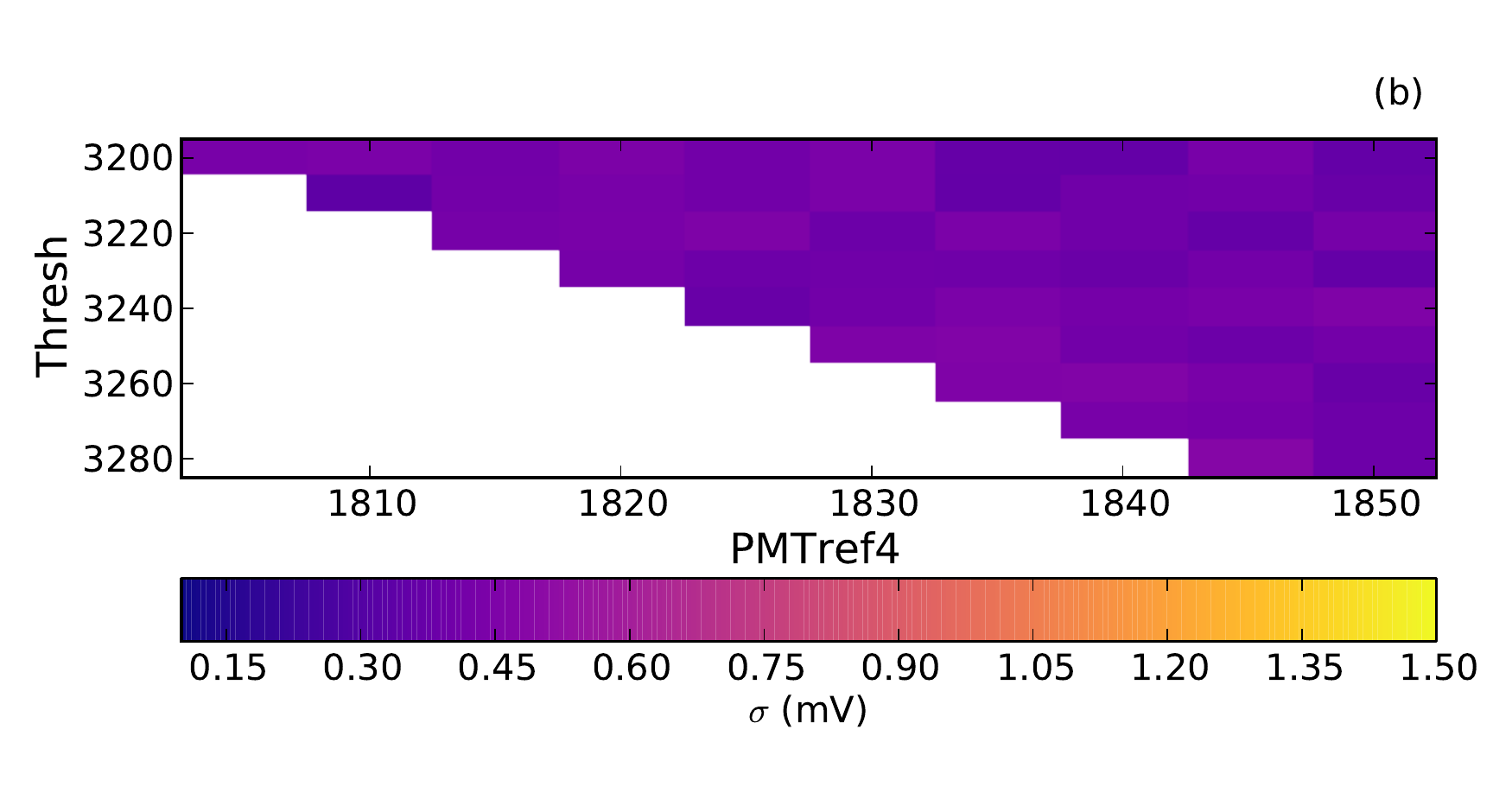}
\caption{Trigger threshold (a) and noise (b) as a function of \tp{PMTref4} and \tp{Thresh} (in DAC counts) with analog sampling disabled. White indicates a 
region of parameter space where the trigger does not function properly.}
\label{trigger-perf-samploff}
\end{figure}

In this configuration, the minimum workable threshold is $\lesssim 5$~mV (1.2 p.e.), and the trigger noise is {\mod ${\lesssim}0.5$~mV} (0.13 p.e.).
Figure ~\ref{figures/trigger-Scurve-samploff.pdf} shows an efficiency curve for sampling disabled for which the threshold was 4.76~mV. Also in this casfe the measured efficiencies are well fit by an $S$ curve.

\begin{figure}
\includegraphics[width=0.5\textwidth]{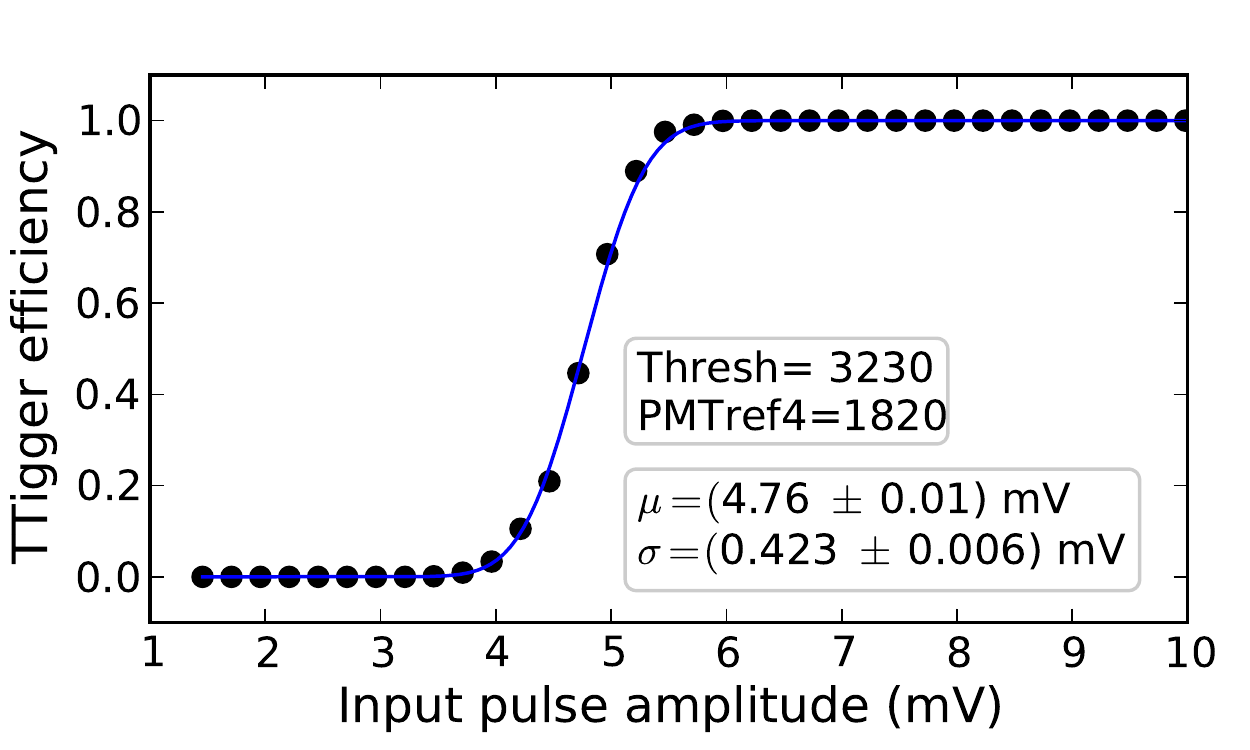}
\caption{Same as Fig.~\ref{trigger-Scurve-normal}, but the signal sampling in the ASIC was disabled. The inset text gives one of the best parameter sets (\tp{Thresh} and \tp{PMTref4}) used for the trigger configuration (see Fig.~\ref{trigger-perf-samploff}).}
  \label{figures/trigger-Scurve-samploff.pdf}
\end{figure}

In conclusion, the performance of the trigger circuit with sampling disabled 
meets the desired {sensitivity and noise level}. Therefore, two options were considered: 1) {\mod reducing the interference on the ASIC between sampling and triggering circuits (by improving isolation and increasing the gain on the trigger path)}, 2) 
separating data and trigger path into two different ASICs. Option 1) was chosen 
as the most cost effective for the design of TARGET~7, {\mod and option 2) for subsequent ASIC pairs}.

\begin{figure}
\includegraphics[width=0.5\textwidth]{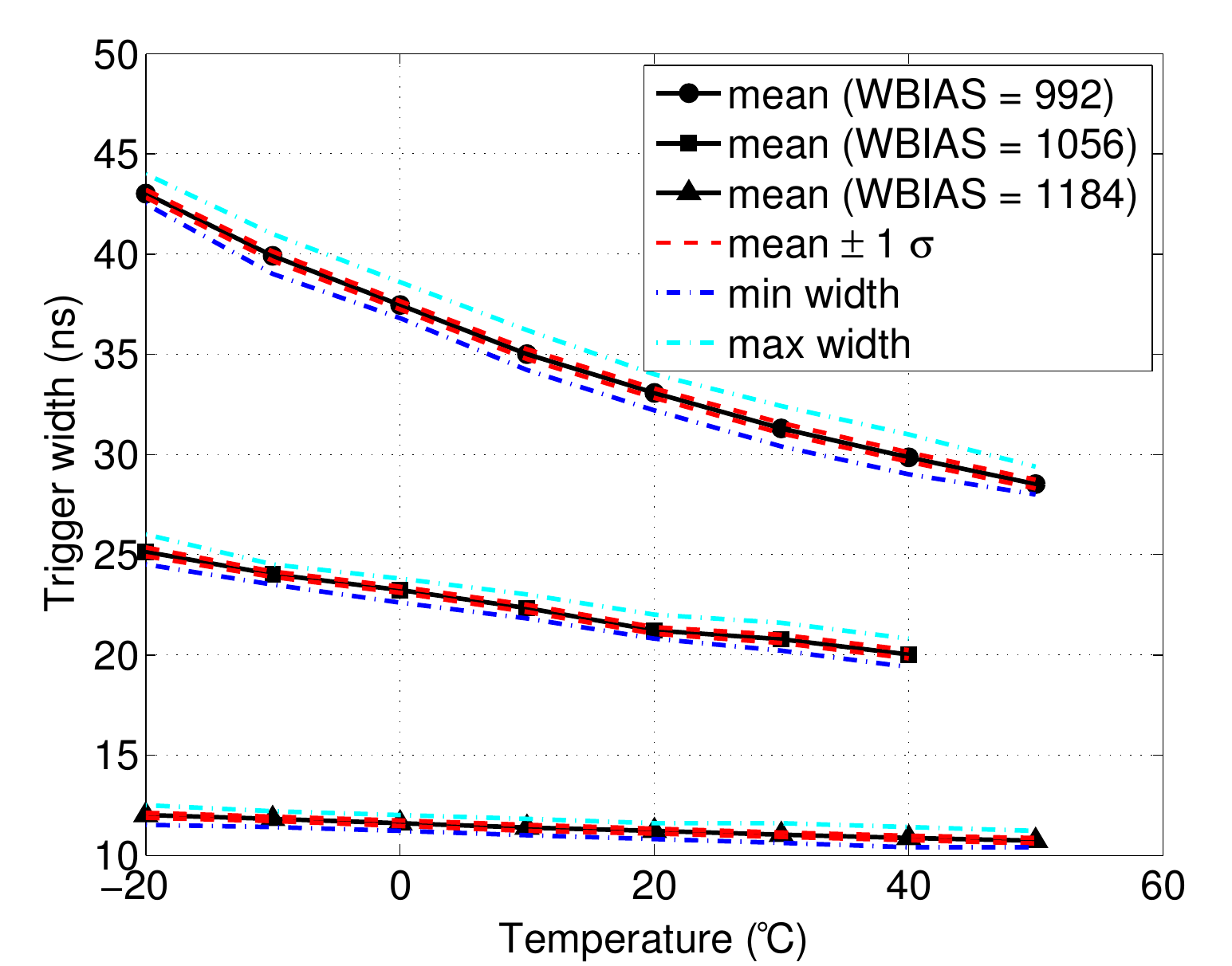}
\caption{Width of the output trigger signals as a function of temperature for different values of the \tp{WBias} control voltage.}
\label{fig:trigvswbias}
\end{figure}

\subsubsection{Trigger output}

The trigger output signal characteristics are summarized in Figure~\ref{fig:trigvswbias}.  The output signal has 2 V amplitude and a width that is tunable using a control voltage (\tp{WBias}).  The chip produces pulses with stable width and amplitude for pulses as narrow as 10~ns.  This width exhibits some temperature dependence, as shown in Figure~\ref{fig:trigvswbias}.  However, the temperature dependence is weak for narrow ($\sim$10~ns) pulses that will be used for most applications.

The maximum {\rev sustained trigger rate achievable without event loss for a waveform length of 2 blocks, i.e., 64 samples,} was measured to be ${\sim}7$~kHz. The limit in the test setup is dominated by the UDP link and a prototype data acquisition software. Nevertheless, the readout rate achieved is {\mod larger than} the rate of ${\sim}600$~Hz expected for {\mod the GCT camera, and that of a few kHz expected for the SCT camera}. 

\section{Front-end electronics module}\label{sec:cameramod}

A front-end electronics module based on TARGET~5 has been designed and produced for a prototype CTA camera, CHEC-M \cite{2013arXiv1307.2807D,2015arXiv150901480D}.  This camera is designed for use with Schwarzschild--Couder small-sized telescopes and features 32 TARGET~5 modules per camera, each reading out a {64-pixel MAPMT \mod (which corresponds to 128 TARGET~5 ASICs per camera)}.  35 {\mod TARGET~5} modules were produced and are currently undergoing commissioning as part of the integrated camera. A module is shown in Figure~\ref{fig:ModulePhoto}. CHEC-M and its front-end electronics system will be described in more detail in subsequent publications.

\begin{figure}
\includegraphics[width=0.5\textwidth]{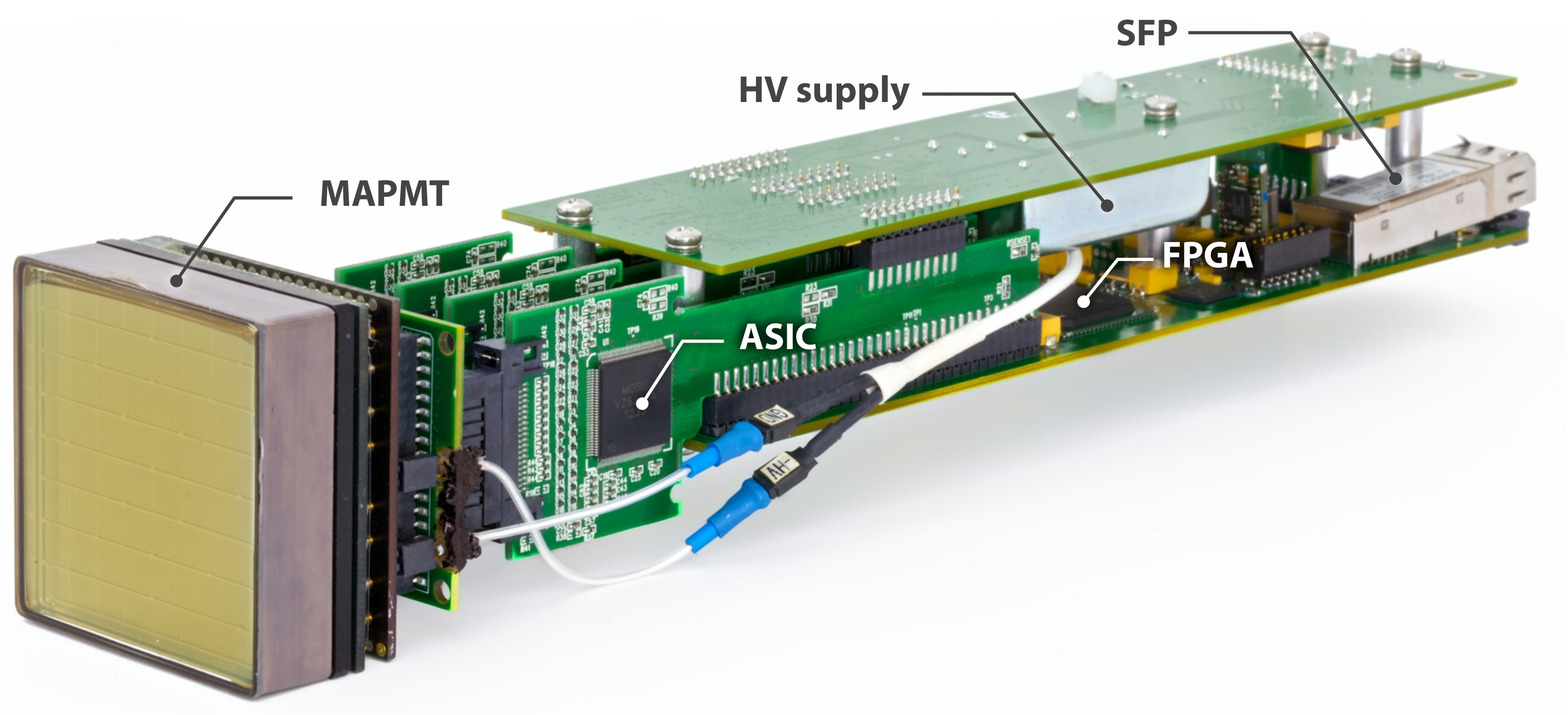}
\caption{TARGET~5 camera module for the CHEC-M prototype CTA camera.  Signals from a single 64-channel MAPMT flow to four TARGET~5 chips through preamplifier boards (not shown in this picture).  The TARGET~5 chips are configured and controlled by a single FPGA. A module comprises four ASIC boards, an FPGA board, and a power board. The module also provides high voltage (HV) to the MAPMT. In the prototype module shown here there is also and SFP optic fiber connector to interface the module with an external computer.}
\label{fig:ModulePhoto}
\end{figure}

\section{Conclusion and outlook}\label{sec:conclusions}

We have developed a new ASIC of the TARGET family designed to read out signals from the photosensors in cameras of very-high-energy gamma-ray telescopes exploiting {\mod time-resolved} imaging of Cherenkov light from air showers. TARGET~5 processes signals from 16 photodetector pixels in parallel both for sampling and digitization and for trigger formation.

Key aspects of the TARGET~5 performance are:
\begin{itemize}
\item sampling frequency tunable between {\rev $< 0.4$~GSa/s and $>1$~GSa/s} (Fig.~\ref{sampling-frequency})
\item a dynamic range on the data path of 1.2 V with {\mod effective dynamic range 11}~bits and DC noise ${\sim}0.6$~mV (Fig.~\ref{temperature})
\item 3-dB bandwidth of 500 MHz (Fig.~\ref{fig:bandwidth})
\item {\rev crosstalk between neighboring channels $<1.3\%$ (Fig.~\ref{crosstalk})}
\item {\mod charge resolution improving from 40\% to $<4\%$ as a function of input charge between 3 p.e. and $>100$~p.e (assuming 4 mV per p.e.)} (Fig.~\ref{charge-linearity} and \ref{charge-resolution})
\item minimum stable trigger threshold of 20 mV (5 p.e.) with trigger noise of 5 mV (1.2 p.e.), mostly limited by {\mod interference between sampling and trigger operations} (Fig.~\ref{trigger-Scurve-normal})
\item minimum stable trigger threshold of $<5$~mV (1.2 p.e.) with trigger noise of 0.5 mV (0.12 p.e.) with sampling disabled (Fig.~\ref{figures/trigger-Scurve-samploff.pdf})
\end{itemize}

TARGET~5 is part of the front-end electronics system of the first GCT camera prototype, also known as CHEC-M \cite{2013arXiv1307.2807D,2015arXiv150901480D}, and is the first ASIC in the TARGET family to be used in an IACT {\mod prototype} which is proposed in the framework of the CTA project.

To meet the performance desired for CTA, further developments are ongoing that are briefly outlined in \cite{tibaldoICRC2015}.

\section*{Acknowledgements}
This research was supported by the SLAC LDRD Grant ``Integrated TeV Gamma-ray Camera Readout System'' and by the Stanford University Dean's Grant ``Cherenkov Telescope Array'', as well as by JSPS KAKENHI Grant Numbers 25610040 and 23244051. A.~O. was supported by a Grant-in-Aid for JSPS Fellows. Additional support to J.~V. and T.~W. was provided by the University of Wisconsin.

We are grateful for many fruitful discussions with members of the CTA consortium, in particular of the GCT and SCT groups. We are especially grateful for assistance in evaluating and optimizing the ASIC performance by Sonia Karkar, and for feedback on the measurements and paper draft from Felix Werner and Scott Wakely.

This paper has gone through internal review by the CTA Consortium.

\bibliographystyle{elsarticle-num}
\bibliography{ref}

\begin{thebibliography}{10}
\expandafter\ifx\csname url\endcsname\relax
  \def\url#1{\texttt{#1}}\fi
\expandafter\ifx\csname urlprefix\endcsname\relax\def\urlprefix{URL }\fi
\expandafter\ifx\csname href\endcsname\relax
  \def\href#1#2{#2} \def\path#1{#1}\fi

\bibitem{2009ARA&A..47..523H}
J.~A. {Hinton}, W.~{Hofmann}, {Teraelectronvolt Astronomy}, \araa 47 (2009)
  523--565.
\newblock \href {http://arxiv.org/abs/1006.5210} {\path{arXiv:1006.5210}},
  \href {http://dx.doi.org/10.1146/annurev-astro-082708-101816}
  {\path{doi:10.1146/annurev-astro-082708-101816}}.

\bibitem{2011ExA....32..193A}
M.~{Actis}, G.~{Agnetta}, F.~{Aharonian}, A.~{Akhperjanian}, J.~{Aleksi{\'c}},
  E.~{Aliu}, D.~{Allan}, I.~{Allekotte}, F.~{Antico}, L.~A. {Antonelli},
  et~al., {Design concepts for the Cherenkov Telescope Array CTA: an advanced
  facility for ground-based high-energy gamma-ray astronomy}, Experimental
  Astronomy 32 (2011) 193--316.
\newblock \href {http://arxiv.org/abs/1008.3703} {\path{arXiv:1008.3703}},
  \href {http://dx.doi.org/10.1007/s10686-011-9247-0}
  {\path{doi:10.1007/s10686-011-9247-0}}.

\bibitem{2007APh....28...10V}
V.~{Vassiliev}, S.~{Fegan}, P.~{Brousseau}, {Wide field aplanatic two-mirror
  telescopes for ground-based {$\gamma$}-ray astronomy}, Astroparticle Physics
  28 (2007) 10--27.
\newblock \href {http://arxiv.org/abs/astro-ph/0612718}
  {\path{arXiv:astro-ph/0612718}}, \href
  {http://dx.doi.org/10.1016/j.astropartphys.2007.04.002}
  {\path{doi:10.1016/j.astropartphys.2007.04.002}}.

\bibitem{Wood201611}
M.~Wood, T.~Jogler, J.~Dumm, S.~Funk, Monte carlo studies of medium-size
  telescope designs for the cherenkov telescope array, Astroparticle Physics 72
  (2016) 11 -- 31.
\newblock \href
  {http://dx.doi.org/http://dx.doi.org/10.1016/j.astropartphys.2015.04.008}
  {\path{doi:http://dx.doi.org/10.1016/j.astropartphys.2015.04.008}}.

\bibitem{2012APh....36..156B}
K.~{Bechtol}, S.~{Funk}, A.~{Okumura}, L.~L. {Ruckman}, A.~{Simons},
  H.~{Tajima}, J.~{Vandenbroucke}, G.~S. {Varner}, {TARGET: A multi-channel
  digitizer chip for very-high-energy gamma-ray telescopes}, Astroparticle
  Physics 36 (2012) 156--165.
\newblock \href {http://arxiv.org/abs/1105.1832} {\path{arXiv:1105.1832}},
  \href {http://dx.doi.org/10.1016/j.astropartphys.2012.05.016}
  {\path{doi:10.1016/j.astropartphys.2012.05.016}}.

\bibitem{2015arXiv150806472M}
T.~{Montaruli}, G.~{Pareschi}, T.~{Greenshaw}, {The small size telescope
  projects for the Cherenkov Telescope Array}, in: Proceedings of the 34th
  International Cosmic Ray Conference, Vol. POS(ICRC2015)1043, 2015.
\newblock \href {http://arxiv.org/abs/1508.06472} {\path{arXiv:1508.06472}}.

\bibitem{2013arXiv1307.2807D}
M.~K. {Daniel}, R.~J. {White}, D.~{Berge}, J.~{Buckley}, P.~M. {Chadwick},
  G.~{Cotter}, S.~{Funk}, T.~{Greenshaw}, N.~{Hidaka}, J.~{Hinton},
  J.~{Lapington}, S.~{Markoff}, P.~{Moore}, S.~{Nolan}, S.~{Ohm}, A.~{Okumura},
  D.~{Ross}, L.~{Sapozhnikov}, J.~{Schmoll}, P.~{Sutcliffe}, J.~{Sykes},
  H.~{Tajima}, G.~S. {Varner}, J.~{Vandenbroucke}, J.~{Vink}, D.~{Williams},
  f.~t. {CTA Consortium}, {A Compact High Energy Camera for the Cherenkov
  Telescope Array}, in: Proceedings of the 33rd International Cosmic Ray
  Conference, 2013.
\newblock \href {http://arxiv.org/abs/1307.2807} {\path{arXiv:1307.2807}}.

\bibitem{2015arXiv150901480D}
A.~{De Franco}, R.~{White}, D.~{Allan}, T.~{Armstrong}, T.~{Ashton},
  A.~{Balzer}, D.~{Berge}, R.~{Bose}, A.~M. {Brown}, J.~{Buckley}, P.~M.
  {Chadwick}, P.~{Cooke}, G.~{Cotter}, M.~K. {Daniel}, S.~{Funk},
  T.~{Greenshaw}, J.~{Hinton}, M.~{Kraus}, J.~{Lapington}, P.~{Molyneux},
  P.~{Moore}, S.~{Nolan}, A.~{Okumura}, D.~{Ross}, C.~{Rulten}, J.~{Schmoll},
  H.~{Schoorlemmer}, M.~{Stephan}, P.~{Sutcliffe}, H.~{Tajima}, J.~{Thornhill},
  L.~{Tibaldo}, G.~{Varner}, J.~{Watson}, A.~{Zink}, {The first GCT camera for
  the Cherenkov Telescope Array}, 2015.
\newblock \href {http://arxiv.org/abs/1509.01480} {\path{arXiv:1509.01480}}.

\bibitem{SCTICRC2015}
A.~N. {Otte}, J.~{Biteau}, H.~{Dickinson}, S.~{Funk}, T.~{Jogler}, C.~A.
  {Johnson}, P.~{Karn}, K.~{Meagher}, H.~{Naoya}, T.~{Nguyen}, A.~{Okumura},
  M.~{Santander}, L.~{Sapozhnikov}, A.~{Stier}, H.~{Tajima}, L.~{Tibaldo},
  J.~{Vandenbroucke}, S.~{Wakely}, A.~{Weinstein}, D.~A. {Williams}, f.~t. {CTA
  Consortium}, {Development of a SiPM Camera for a Schwarzschild-Couder
  Cherenkov Telescope for the Cherenkov Telescope Array}, in: Proceedings of
  the 34th International Cosmic Ray Conference, Vol. POS(ICRC2015)932, 2015.
\newblock \href {http://arxiv.org/abs/1509.02345} {\path{arXiv:1509.02345}}.

\bibitem{Varner:2007zz}
G.~Varner, L.~Ruckman, P.~Gorham, J.~Nam, R.~Nichol, et~al., {The large analog
  bandwidth recorder and digitizer with ordered readout (LABRADOR) ASIC},
  Nucl.Instrum.Meth. A583 (2007) 447--460.
\newblock \href {http://arxiv.org/abs/physics/0509023}
  {\path{arXiv:physics/0509023}}, \href
  {http://dx.doi.org/10.1016/j.nima.2007.09.013}
  {\path{doi:10.1016/j.nima.2007.09.013}}.

\bibitem{2012JInst...7.8021C}
D.~{Casadei}, {Estimating the selection efficiency}, Journal of Instrumentation
  7 (2012) 8021.
\newblock \href {http://arxiv.org/abs/0908.0130} {\path{arXiv:0908.0130}},
  \href {http://dx.doi.org/10.1088/1748-0221/7/08/P08021}
  {\path{doi:10.1088/1748-0221/7/08/P08021}}.

\bibitem{tibaldoICRC2015}
L.~{Tibaldo}, J.~A. {Vandenbroucke}, A.~M. {Albert}, et~al., in: Proceedings of
  the 34th International Cosmic Ray Conference, Vol. POS(ICRC2015)932, 2015.
\newblock \href {http://arxiv.org/abs/1508.06296} {\path{arXiv:1508.06296}}.

\end{thebibliography}







\end{document}